# Leveraging Large Language Models to Democratize Access to Costly Datasets for Academic Research

September 2025

Julian Junyan Wang
University College, University of Oxford
julian.wang@univ.ox.ac.uk

Victor Xiaoqi Wang
College of Business, California State University Long Beach
victor.wang@csulb.edu

**Abstract**: Unequal access to costly datasets essential for empirical research has long hindered researchers from disadvantaged institutions, limiting their ability to contribute to their fields and advance their careers. Recent breakthroughs in Large Language Models (LLMs) have the potential to democratize data access by automating data collection from unstructured sources. We develop and evaluate a novel methodology using GPT-4o-mini within a Retrieval-Augmented Generation (RAG) framework to collect data from corporate disclosures. Our approach achieves human-level accuracy in collecting CEO pay ratios from approximately 10,000 proxy statements and Critical Audit Matters (CAMs) from more than 12,000 annual reports, with LLM processing times of 9 and 40 minutes respectively, each at a cost under US $10. This stands in stark contrast to the hundreds of hours needed for manual collection or the thousands of dollars required for commercial database subscriptions. To foster a more inclusive research community by empowering researchers with limited resources to explore new avenues of inquiry, we share our methodology and the resulting datasets.

---

We thank Prithviraju Venkataraman for excellent research assistance.

## 1. Introduction

In the realm of academia, the adage "publish or perish" has long been a guiding principle, highlighting the critical importance of research output in scholarly careers. The pressure to publish has intensified in recent decades, as scholarly output now serves as the primary metric for assessing research excellence, advancing academic careers, and establishing institutional rankings (Swanson, 2004; van Dalen & Henkens, 2012). This heightened emphasis on research output has created a highly competitive academic environment, where the ability to conduct and disseminate impactful research is paramount.

Yet this publication-centric paradigm, while ostensibly meritocratic, has inadvertently fostered a landscape of inequality within academia. Well-resourced institutions, with their access to cutting-edge tools, comprehensive databases, and ample research support, stand at a significant advantage. In contrast, researchers at less affluent institutions often find themselves navigating a treacherous path, their scholarly ambitions hampered by limited access to essential resources, data, and infrastructure. Such disparities not only impede individual career progression but also threaten to homogenize the pool of contributors to academic knowledge, potentially stifling the diversity of perspectives vital for robust intellectual discourse and innovation.

Recent breakthroughs in Generative AI (GenAI) and Large Language Models (LLMs), however, offer a potential solution. These technologies have demonstrated unprecedented capabilities in processing unstructured text, automating complex analytical tasks, extracting information from diverse document formats (Levich & Knust, 2025; H. Li et al., 2024), and generating deeper insights from these documents (Cong & Yang, 2024; Jha et al., 2024, 2025). This raises a critical question: Could these emerging technologies serve as equalizing forces by democratizing access to the costly datasets that have become essential for academic research?

This question is particularly pressing in business disciplines like accounting and finance, where the increasing prevalence of empirical research has fundamentally transformed scholarly requirements. Studies now rely heavily on datasets, with research using more datasets being more likely to be published (Berninger et al., 2022; Dai et al., 2023). Publishing novel insights increasingly necessitates unique datasets, which can be challenging and expensive to acquire, especially when data is not commercially available or has only recently emerged from regulatory changes or advances in technology. Researchers from well-funded institutions can obtain these datasets through internal resources or commercial providers, while those at less privileged institutions struggle with acquisition costs—whether through expensive subscriptions or labor-intensive manual collection. This disparity has created a formidable barrier to entry for many aspiring researchers, reinforcing an increasingly homogeneous academic landscape where research perspectives and insights are predominantly shaped by a small number of well-resourced institutions. The resulting lack of diversity in the research community may suppress valuable insights from talented but resource-constrained researchers, limiting scientific progress and innovation.

To address these data access challenges, many prior studies in accounting, finance and other business disciplines have attempted to construct novel datasets by exploring data from unstructured sources, including regulated filings and other corporate documents. These studies primarily relied on rule-based methods to extract entire sections (Bao & Datta, 2014; Dyer et al., 2017; F. Li, 2010; Muslu et al., 2014). However, inconsistent formatting across company documents poses significant challenges for these approaches (Bao & Datta, 2014). Extracting specific information within sections proves even more difficult, leading some recent studies to resort to manual data collection for their research projects on emerging issues (e.g., Bourveau et al., 2025; Demers et al., 2024a).

While traditional approaches have struggled with these challenges, GenAI and LLMs offer new possibilities for automating many routine tasks and transforming how researchers conduct research in accounting, finance, and related fields (Brown et al., 2025; de Kok, 2023; Dong et al., 2024; Dowling & Lucey, 2023; Giesecke, 2024; Korinek, 2023; Wang & Wang, 2025). Building on these technological capabilities, this study explores GenAI's potential to democratize academic research, particularly in quantitative fields like accounting and finance, by equalizing access to costly datasets. We posit that GenAI can transform academic research by broadening participation, expanding the pool of researchers capable of conducting quantitative studies, diversifying the range of topics investigated, and increasing the geographical scope of research. Furthermore, by enabling efficient data collection and analysis, GenAI may allow researchers to redirect their efforts toward more complex and creative aspects of their work (Filetti et al., 2024; H. Li et al., 2024). This study specifically examines automatic data collection using these latest technologies.

To evaluate the potential of LLMs for democratizing access to costly datasets, we focus on two specific types of data from corporate disclosures: CEO pay ratio disclosures and Critical Audit Matters (CAMs). The former is quantitative, while the latter is qualitative, representing the two major types of data utilized in empirical research. Both present ideal test cases because their presentation is unstructured and varies widely in formatting across documents, making them challenging for automatic extraction using traditional methods. These data sources, which have emerged from recent regulatory changes, present numerous research opportunities. However, until now, their utilization has been largely confined to well-funded institutions or those willing to invest substantial time in manual data collection, limiting the scope and diversity of research in these areas.

We employ GPT-4o-mini, a capable and cost-effective LLM from OpenAI, combined with regular expressions to develop a novel methodology for extracting targeted information from complex corporate filings. Built on Retrieval Augmented Generation (RAG) (Lewis et al., 2021), our methodology first retrieves relevant passages from a large corpus and uses them to condition the language model for more accurate output. Through careful prompt engineering and iterative refinement, we guide the LLM to handle various disclosure formats**,** enhancing efficiency and accuracy while reducing processing time and costs.

The results of our large-scale experiments demonstrate both the efficacy and economic viability of our approach. We successfully collect CEO pay ratio data from nearly 10,000 proxy statements in just 9 minutes at a cost of approximately US $7, achieving an accuracy rate exceeding 99%. Similarly, we extract CAMs from more than 12,000 annual reports in around 40 minutes for approximately US $8, yielding an accuracy rate of 98-99% when validated against verified samples. These results are particularly striking when compared to manual data collection or commercial subscriptions, which can take hundreds of hours or cost thousands of dollars. Such extraordinary efficiency underscores the transformative potential of LLMs in reshaping the landscape of data collection.

Our study makes several contributions that span academic research, practical applications, and policy implications. First, we demonstrate that LLMs can democratize access to costly datasets by extracting information from unstructured documents at minimal cost. Our methodology's success with complex, heterogeneous document formats featuring diverse narratives and inconsistent structures suggests its broad applicability across disciplines. This breakthrough is particularly significant for researchers at resource-constrained institutions who have been historically excluded from data-intensive research due to prohibitive costs.

Second, we provide comprehensive methodological guidance through detailed documentation that serves as a practical roadmap for implementation. We offer step-by-step instructions and insights on data preparation, prompt engineering, and API utilization, facilitating adoption of these techniques across the research community. This transparency promotes reproducibility and enables researchers to adapt our approach to their specific data needs.

Third, we contribute directly to the research community by making publicly available two valuable datasets on CEO pay ratios and CAMs, both of which emerge from recent regulatory changes. By sharing these datasets, we join recent initiatives (Bergeaud & Verluise, 2024; deHaan et al., 2024; Demers et al., 2024b) that enable exploration of new questions in executive compensation, corporate governance, and financial reporting through open access to data (Andreoli-Versbach & Mueller-Langer, 2014).

Fourth, our findings extend beyond academia to offer substantial practical and policy value. For market participants such as investors and analysts, our approach reduces data acquisition costs and enhances decision-making capabilities, potentially improving capital market efficiency alongside other AI-driven advances in financial services (Huang et al., 2025; Tan et al., 2025). For regulatory bodies, our methodology enables a paradigm shift from sample-based periodic reviews (Bozanic et al., 2017) to population-wide, real-time compliance monitoring using AI tools at a fraction of traditional costs. More broadly, this work demonstrates how the research community can embrace GenAI as a general method of invention (Bianchini et al., 2022), unlocking previously inaccessible data sources and enabling research questions that were computationally or financially prohibitive (Goos & Savona, 2024). Such advances must, however, adhere to emerging ethical frameworks for sustainable AI deployment (Cumming et al., 2024).

## 2. Background and literature review

### 2.1 Growing importance of data in academic research

In business fields like accounting and finance, the type of research conducted in recent decades has become increasingly empirical and quantitative. Dai *et al*. (2023) conduct an analysis of 52,497 papers posted in the Financial Economics Network (FEN) of the Social Science Research Network (SSRN) from 2001 to 2019, finding that the proportion of empirical research has increased from 68 percent in 2001 to 85 percent in 2019. This finding is consistent with Berninger *et al*. (2022), who document that the share of empirical contributions to finance journals grew from 70 percent in 2000 to almost 90 percent in 2016.

This trend also parallels the pivot from theoretical research to empirical research in the field of economics (Angrist et al., 2020; Hamermesh, 2018). Moreover, the current rise in empirical research in business fields is merely a continuation of a trend that began in the last century. For example, in the *Journal of Financial Economics*, 59 percent of articles were theoretical and only 39 percent were empirical over 1974 to 1979 (Schwert, 2021). However, there has been a radical reversal with 88 percent of papers being empirical over 2010 to 2020 with only 12 percent being theoretical. Kim *et al.* (2006) find a similarly drastic change in 41 finance and economics journals with 77 percent of the most cited papers being theoretical in the 1970s and only 11 percent being theoretical in 2000.

This rise in empirical research in business and other fields is accompanied by an increasing dependence on databases. Dai *et al.* (2023) find that the average number of databases per empirical article has increased from 2.89 to 4.66 between 2001 and 2019. Berninger *et al*. (2022) equally observe an increase from two to more than 3.5 databases used per article, which they partially attribute to growing pressure to use more control variables and robustness checks. According to them, one database does not provide sufficient data to gain insights that warrant publication, leading to more databases being required to address meaningful research questions. Dai *et al*. (2023) demonstrate that this pressure to use more databases is not misplaced as a one standard deviation increase in the number of databases used in a study corresponds to a 26 percent higher likelihood of publication. To produce quality research today, researchers need comprehensive data to align with the increasingly empirical and quantitative nature of the business fields.

Moreover, as common datasets like Compustat and CRSP have been extensively used in accounting and finance research, it is almost impossible to publish novel insights in top journals relying solely on such datasets. Successful publication in premier outlets often hinges on utilizing unique and novel datasets that provide fresh perspectives on and insights into previously unresolved research questions. However, acquiring such datasets can be challenging and costly. In some cases, commercial data providers offer access to these datasets, but often at high subscription fees. Despite the substantial cost, reliance on data providers has become essential in many instances, as publicly accessible raw data is typically unstructured and often decentralized, making efficient use of the data particularly burdensome.

In certain situations, data only recently becomes available through regulatory changes or technological advancements and lacks commercial distribution in user-friendly formats. Additionally, some datasets serve niche interests, offering little economic incentive for providers to collect and sell them given limited demand. Under these circumstances, researchers must manually gather and curate the data themselves, creating substantial additional work.

As the demand for data-driven insights in accounting, finance, and many other fields continues to grow, the importance of novel datasets for publishing in top journals is expected to increase further. Consequently, researchers who can identify, collect, and analyze unique data sources are likely to have a competitive advantage in producing high-impact research that pushes the boundaries of current knowledge in their fields.

### 2.2 Limited access to data at disadvantaged institutions

The growing importance of data in research has highlighted the unfortunate reality that access to data is unequal due to financial barriers. Borgman (2015) borrows from Anderson (2004) to suggest a "long tail" distribution of data access where there exists a small number of well-funded research teams working with large volumes of data, some teams working with almost no data, and most teams falling in between. Berninger *et al*. (2022) demonstrate this unequal data access in financial research empirically. They show that researchers affiliated with top business schools tend to use user-friendly datasets that are more expensive, whereas researchers from lower-ranking business schools rely more on less expensive, often harder-to-use data sources, which may primarily serve business professionals rather than academics.

This reality raises significant concerns about equity and access in academic research, particularly for scholars at smaller institutions with limited funding. These researchers often face insurmountable obstacles in acquiring or creating novel datasets due to financial constraints, lack of research assistance, and limited technological infrastructure. Unlike their counterparts at well-funded universities, faculty at smaller institutions typically juggle heavier teaching loads, leaving less time for the labor-intensive tasks of data collection and curation. The inability to access or create novel datasets puts these researchers at a significant disadvantage when competing for publication in top journals, potentially creating a self-reinforcing cycle where they struggle to build the publication record necessary to advance their careers.

As data becomes increasingly crucial for research in accounting, finance, and many other business disciplines, addressing this inequality in data access will be essential to ensure that all researchers have the opportunity to conduct impactful and innovative studies. Without equal access to comprehensive and user-friendly datasets, researchers at institutions with limited resources may struggle to contribute to the advancement of their fields, potentially limiting the diversity and quality of research produced in these disciplines.

### 2.3 Impact on research productivity

The literature on research productivity identifies various determinants at the individual, institutional, and national levels (Beaudry & Allaoui, 2012; Dundar & Lewis, 1998; Heng et al., 2021; Simisaye, 2019; Wanner et al., 1981). Availability of funding is a crucial institutional factor that can increase research productivity by enabling academics to attend conferences, publish work, and acquire reference materials (Bland & Ruffin, 1992; Lertputtarak, 2008). Research funds can also increase productivity by providing access to research assistants. Dundar and Lewis (1998) find that research-doctorate programs with greater financial support and a greater percentage of graduate students serving as research assistants saw greater departmental research productivity. Research assistants can gather data from decentralized and unstructured sources, serving as a substitute for expensive databases. Conversely, management faculty at business schools with higher teaching loads, characteristic of less-funded institutions, have lower research productivity (K. Kim & Choi, 2017). These findings emphasize the importance of addressing unequal access to data and research resources across institutions.

While co-authoring with researchers from institutions with data access is a potential solution, it presents several challenges. First, researchers from institutions with limited resources may struggle to find suitable collaborators with access to required data. This can be due to a lack of established networks or the reluctance of researchers from well-funded institutions to collaborate with those from less-resourced ones. Second, even when collaborations are established, researchers without direct data access may have less control over the research process and depend on collaborators for data-related tasks. Such dependency can create power imbalances and impede researchers' ability to fully explore their research questions or preferred methodological approaches. Third, relying on collaborations with data-rich institutions may limit the diversity of research perspectives and questions explored, due to the less control over the research process and the difference in targeted journals.

### 2.4 AI and research productivity

Given the significant impact of financial barriers and unequal access to data on research productivity, it is crucial to explore potential solutions to level the playing field. The critical issue is whether digital tools, especially GenAI, can "level the playing field" and contribute to a more equitable research landscape. Indeed, many researchers currently believe that GenAI can increase researchers' productivity and contribute to a "democratization" of academic research. In a survey of 1,600 researchers, the most popular answer to a question on the biggest benefit of GenAI in research was to support researchers who do not speak English as their first language (Van Noorden & Perkel, 2023). This suggests that GenAI could help reduce language barriers and enable a more diverse group of researchers to contribute to the global scientific community.

In the context of quantitative research, Filetti et al. (2024) suggest that GenAI can streamline the research process by automating menial tasks such as data cleaning and normalization. By reducing the time and effort required for these tasks, GenAI could allow researchers to focus on more complex and value-added aspects of their work, potentially leading to increased research productivity. Already, there are examples or evidence of how researchers may use GenAI to replace or enhance certain tasks. For instance, Dowling and Lucey (2023) demonstrate that ChatGPT can significantly assist with finance research, excelling in idea generation and data identification. Similarly, Korinek (2023) explores how LLMs such as ChatGPT can assist economists in various aspects of the research process, from ideation and writing to data analysis, coding, and mathematical derivations.

The ability of new technologies to revolutionize academic research and "level the playing field" is not new. For example, the development of communication technologies enabled the possibility of greater collaboration (e.g. co-authorship) which particularly benefitted middle-tier universities and weakened the competitive edge of elite universities (Agrawal & Goldfarb, 2008; E. H. Kim et al., 2009). This instance highlights how technological advancements can disrupt traditional power dynamics in academia and create a more equitable research landscape.

It is important to note, though, that the case of communication technology specifically affected the logistics of conducting research and not the research itself. In contrast, recent technological advances such as machine learning and GenAI have enabled researchers to be more efficient in conducting various aspects of research, leading to savings in both time and financial costs (Dowling & Lucey, 2023; Przybyła et al., 2018). These technologies have the potential to directly impact the research process by automating tasks, extracting insights from large volumes of data, and supporting researchers in their analysis and interpretation of findings.

In this study, we examine whether GenAI has the potential to democratize research, specifically by investigating its ability to democratize or equalize access to expensive datasets, which are essential for conducting quantitative research, a dominant type of research in many disciplines. The term "democratization" has frequently permeated discussions of GenAI, and it is important to clarify that democratization does not necessarily mean "leveling the playing field." Rather, the reverse is true. Etymologically, "democracy" refers to giving power to the people, and "democratization," as applied to academic research, would reasonably mean broadening academic research to include a larger population. "Leveling the playing field" is, therefore, one way of achieving "democratization."

The use of GenAI to enable researchers to quickly collect data at minimum cost could democratize academic research in three ways: broadening the group of researchers able to perform quantitative research, broadening the range of topics studied quantitatively, and broadening the geographic range of countries studied. Firstly, GenAI could empower researchers who were previously unable to conduct quantitative research due to financial barriers limiting their access to data. The latest technology has the potential to allow researchers to collect and structure publicly available data that exists in unstructured formats. For instance, OpenAI's most recent version of a cost-effective yet highly powerful model ("GPT-4o-mini") costs as little as US$0.15 per million input tokens, making large-scale data collection financially accessible to a wide range of researchers.

Secondly, using GenAI to collect data could broaden the range of topics studied quantitatively. Borgman (2015) remarks that large volumes of data (i.e. those contained in large datasets) tend to lack variety and are instead "homogenous in content and structure." Large data providers must standardize data formats for consistency due to their broad user base. For instance, they may standardize the coding of variables, potentially suppressing alternative interpretations of the same qualitative information. Consequently, researchers have limited flexibility in the topics they can explore or construct variables that better address their research questions. However, GenAI enables researchers to collect their own data, granting greater control over measurement choices, research designs, and results interpretations. This will allow researchers to study topics that may have previously lacked the broad appeal necessary for attention from large data providers.

Thirdly, GenAI can broaden the geographic range of countries studied quantitatively. Karolyi (2016) exposes an "academic home bias puzzle" where there is a strong US-centric tilt in financial research and many other fields. He finds that only 16 percent of all empirical publications in the top four finance journals use non-American data, while many other countries are underrepresented. This bias could be partly attributed to the large size of the American stock market, which enables a maximized sample size for quantitative research (Berninger et al., 2022). However, Karolyi (2016) also points to poor data access as a key contributor. Moreover, Karolyi (2016) notes that "enterprising scholars could dig up sources for successful outcomes," although this often incurs financial costs, which is a barrier to many researchers. As such, GenAI can increase the range of countries studied quantitatively by enabling researchers to cheaply collect data for countries or regions that have previously been overlooked. Therefore, in these three ways, GenAI has the potential to contribute to the democratization of academic research.

### 2.5 Using GenAI to collect data

Specifically, this study focuses on the potential of LLMs for research democratization by investigating their capability for automating data collection from unstructured sources. Primarily relying on rule-based methods, many prior studies have extracted full sections of text from various types of corporate documents, for example, regulated filings (Bao & Datta, 2014; Dyer et al., 2017; e.g., F. Li, 2010; Muslu et al., 2014). Company-specific variations in formatting pose significant challenges for extracting data from these documents. While researchers have developed various approaches to address these challenges (e.g., El-Haj et al., 2020), the inconsistency across documents continues to complicate automated extraction efforts. Machine learning (ML) techniques offer a potential solution to this problem. However, the effectiveness of traditional ML-based methods, which often require model fine-tuning, remains unclear and could significantly increase technical difficulty and costs.

Recently, Li et al. (2024) and Levich and Knust (2025) explore the potential of GenAI to collect tabulated data from PDF documents using Large Language Models (LLMs) on a small sample of documents. Our study extends this emerging line of research in three important ways. First, we focus on both quantitative and qualitative data, including untabulated information, which is an underexplored area. Second, we conduct large-scale experiments to systematically identify challenges in processing extensive datasets. Third, we implement an RAG framework that optimizes processing time and costs when handling large volumes of text.

Our methodology builds on Retrieval Augmented Generation (RAG), a technique introduced by Lewis et al. (2021) that enhances LLM performance by combining advanced language modelling with precise information retrieval. In our implementation, we first extract relevant passages from lengthy documents, each of which contains tens of thousands of words, and then prompt the model to process these extracted passages with strict adherence to the original text. This RAG-based approach offers several advantages: (1) Cost-effectiveness. By targeting specific relevant sections, we minimize the amount of text fed into the LLM, significantly reducing the number of tokens processed and resulting in lower computational costs associated with LLM usage. (2) Processing efficiency. By minimizing extraneous text, our selective retrieval approach significantly reduces overall task completion time. (3) Enhanced accuracy. By providing focused, relevant context, we reduce the likelihood of model hallucinations (i.e., generating incorrect or nonsensical information) and ensure that the LLM's responses are grounded in accurate, context-specific information.

## 3. Data sources and experimental tasks

While acknowledging the critique of US-centric studies, we strategically focus on the US Securities and Exchange Commission (SEC) filings for several reasons. The SEC's EDGAR system hosts over 20 million filings and grows by 3,500 filings daily, providing an extensive dataset ideal for testing LLM performance on large samples.[1] Moreover, a large portion of these filings come from foreign registrants, providing substantial international representation.

Our methodology has wide potential across various jurisdictions and is not limited to SEC filings. The use of US data serves as a proof of concept, demonstrating GenAI's potential in processing large volumes of unstructured text that vary in presentation form and formatting. The task complexity we tackle in this study, rather than the specific format or regulatory framework, highlights the generalizability of our approach to other types of documents. The insights from this study are readily adaptable to other regulatory contexts, and the framework we develop and use can be tailored to various types of documents.

For our tests, we focus on data that results from two recent regulations: the CEO pay ratio disclosure and the Critical Audit Matter (CAM) disclosure in the US. As mandated by the Dodd-Frank Act, public companies are required to disclose the ratio of the CEO's annual total compensation to the median compensation of all other employees. The SEC adopted the final rule implementing the pay ratio disclosure requirement in August 2015, and it became effective for fiscal years beginning on or after January 1, 2017. The pay ratio disclosure has attracted significant attention from researchers (Boo et al., 2024; Boone et al., 2024; e.g., Cheng & Zhang, 2023)), as it offers new insights into income inequality within firms and the potential effects of pay disparities on employee morale, productivity, and firm performance.

[1] https://www.sec.gov/submit-filings/about-edgar

CAMs are significant issues that auditors communicate to the audit committee, which are required to be disclosed in the auditor's report under the new auditing standard AS 3101. The Public Company Accounting Oversight Board (PCAOB) adopted AS 3101 in 2017, and it became effective for audits of fiscal years ending on or after June 30, 2019, for largest public companies in the US, and December 15, 2020, for all other companies to which the requirement applies. The disclosure of CAMs provides valuable insights into the most significant risks and uncertainties faced by companies, as well as the auditor's perspective on these issues. Early studies on CAMs have provided valuable findings (e.g., Bentley et al., 2021; Beyer et al., 2024; Burke et al., 2023; Klevak et al., 2023). These studies primarily come from institutions with the financial resources to purchase data from providers, which collect the data from 10-K filings.[2]

We have chosen these two types of data for several reasons. First, these disclosures come in a wide variety of formats and are not tagged using XBRL, making them challenging to collect through traditional automated methods. The language and terminology used in these disclosures can also vary significantly, further complicating the use of automated collection methods. As a result, manual collection has been necessary to accurately gather this data before the recent breakthrough in GenAI.

Second, these two data types reflect the challenges faced by researchers across many fields. Pay ratio disclosures are not readily available from commercial data providers. CAM disclosures, while available from commercial providers, require substantial subscription fees that can be prohibitive for some institutions. These datasets illustrate data accessibility challenges for resource-limited institutions, as both require either manual collection or significant financial expenditure. Furthermore, pay ratio disclosures involve quantitative data, whereas CAMs represent qualitative data. By examining both data types, we test LLMs' ability to handle quantitative and qualitative information, providing a comprehensive assessment of their capabilities.

[2] 10-K filings are comprehensive annual reports that publicly traded companies in the United States are required to file with the Securities and Exchange Commission (SEC).

Third, the data is embedded in large documents, presenting another challenge. In our sample, an average 10-K filing contains over 65,000 words, and an average proxy statement contains nearly 40,000 words. Presenting entire documents to LLMs may not be feasible due to the limited context windows of LLMs or the prohibitive computational cost. To address this issue, we apply Retrieval Augmented Generation (RAG), a relatively new technique that significantly enhances the accuracy and cost-effectiveness of data collection by focusing on the relevant sections of documents.

Fourth, these data emerge from recent regulations that offer abundant research opportunities. As these regulations are relatively new, their impacts on corporate governance, executive compensation, and financial reporting remain underexplored. By documenting our data collection process and sharing these datasets, we aim to democratize research access. Making these resources available to researchers with limited financial means enables a broader range of institutions and scholars to study these important topics. This fosters a more diverse and inclusive research community, bringing about varied perspectives on and insights into the study of these regulatory changes.

## 4. Methodology

Extracting data from CEO pay ratio disclosures presents challenges due to varying formats and narratives across companies, as illustrated in Appendix A. The inconsistent formatting makes it difficult for rule-based methods and traditional ML-based algorithms to accurately identify and extract relevant data. Similarly, Critical Audit Matters (CAMs) in auditors' reports from 10-K filings differ significantly across companies, as shown in Appendix B. The varied structure, formatting, and language patterns complicate consistent data extraction using traditional methods.

To address these challenges, we leverage Large Language Models (LLMs) and data processing techniques within a Retrieval-Augmented Generation (RAG) framework. Specifically, we use the “gpt-4o-mini” model via the OpenAI API. Providing an optimal balance of performance and cost-effectiveness, this model, released on July 18, 2024, features a 128K context window and an output capacity of 16,384 tokens. Moreover, this model charges only USD 0.15 per million input tokens and USD 0.6 per million output tokens.

Our methodology comprises several key steps, including downloading and parsing relevant filings, developing regular expressions to extract specific sections, performing prompt engineering to ensure accurate and consistent data extraction from LLMs, and querying the API with carefully crafted prompts and input text extracts. We employ an iterative process for prompt engineering, starting with simple prompts and gradually refining them based on the model’s performance on a small sample of extracts. The final prompts provide clear and detailed instructions to the model, guiding it to identify, collect, and structure the required information while minimizing the risk of hallucination. Please refer to Appendix C for full details of the entire process.

## 5. Experimental results

### 5.1 Sample selection

Our sample is limited to Compustat Execucomp companies, as pay ratio disclosure studies typically require CEO attributes and other variables from this database. Our final sample consists of 9,865 proxy statements containing pay ratio disclosures (filed 2018-2023) and 12,499 10-K forms containing CAM disclosures (filed 2019-2023). These date ranges begin with the first year each disclosure type became mandatory: 2018 for pay ratios and 2019 for CAMs. The sample selection process is summarized in Panel B of Table 1.

### 5.2 Results for CEO pay ratio data

#### 5.2.1 Results of initial passage extraction

In our RAG framework, the first crucial step involves extracting relevant passages from source documents. These extracts are then provided to the chosen LLM for data collection. To extract pay ratio disclosures from proxy statements, we employ a systematic approach to extract relevant content. For most filings, we are able to programmatically identify pay ratio disclosure headings, allowing for a single, comprehensive extract. In cases where such headings are not readily identifiable, we rely on references to median employee pay, sometimes resulting in multiple extracts per file to ensure the capturing of the pay ratio data. Table 2 presents the distribution of extracts across our sample filings. Panel A shows that most files (73.90%, n=7,290) yield a single extract. From our total sample of 9,865 proxy statements, we obtain 13,960 extracts, averaging 1.41 extracts per file. For files with multiple extracts, we feed all of them to the LLM to ensure that the relevant data is captured.

### 5.2.2 Input tokens, processing time, and cost

We process one extract per API request, as larger batch sizes risk cross-contamination of data across extracts. The prompt shown in Fig. 1 consists of 1,114 tokens, and each extract contains 1,821 tokens on average. The total input tokens are 40.97M: 15.55M from prompts (1,114 tokens × 13,960 requests) and 25.42M from extracts (1,821 tokens × 13,960 extracts).

Our implementation processes these 13,960 extracts through individual API requests, incorporating automated error handling and retry mechanisms. The "gpt-4o-mini" model successfully processed all extracts in approximately nine minutes, incurring a total cost of $7 in API fees. For comparison, manual collection, estimated at three minutes per filing for a total of 9,865 filings, would require approximately 493 hours. This translates to 62 working days, assuming an eight-hour working day, or three calendar months when holidays are considered. At a rate of US $10 per hour, manual collection would cost approximately $5,000. Our LLM-based method demonstrates a significant reduction in time and cost, transforming months of manual labor into mere minutes of computational time at just 0.14% of the estimated manual labor cost.

It is worth noting that our approach scales efficiently to larger samples, costing approximately $0.50 per thousand extracts ($7 / 13,960 × 1,000). For each additional year, with around 1,500 filings, the cost increases by only about one dollar. Furthermore, this method can be easily adapted to extract additional information (e.g., explanations of how median employee pay is determined) from the same documents at minimal extra cost, simply by adjusting the prompt.

### 5.2.3 Accuracy

As shown in Panel A of Table 3, out of 9,865 proxy statements, the model successfully collected CEO compensation from 9,756 statements (98.90%), median employee pay data from 9,839 statements (99.74%), and pay ratio figures from 9,849 statements (99.84%). These remarkably high collection rates across all three metrics, with missing percentages ranging from just 0.16% to 1.10%, underscore the model's reliability and robust performance in handling diverse data presentations within proxy statements. It is worth mentioning that the missing elements do not necessarily mean that the model missed them. In some cases, the extracts provided to the model do not contain the relevant information.

We rigorously evaluate our approach by focusing on accuracy rather than metrics such as recall, precision, or F1 score. This emphasis on accuracy suits our task design: instead of performing binary or multi-class classification, we collect specific numerical values from text. Our methodology employs RAG to identify and process only relevant text segments containing pay ratio information, minimizing processing time and costs by reducing LLM use on irrelevant text. Given this setup, accuracy becomes the most meaningful metric. Both precision and recall should theoretically align with accuracy, as the LLM either correctly extracts the values or not. This alignment occurs because our task involves accurately gathering specific numerical values from provided text sections, not identifying all possible mentions (recall), or avoiding false positives (precision).

First, we assess the internal consistency of the collected data, ensuring that the collected pay ratio is equal to the ratio calculated between the collected CEO compensation and median employee pay. Second, for observations where we are unable to compare the collected ratio against the calculated ratio due to missing data, we manually verify the accuracy of these observations.[3] Third, we compare a sample of approximately 2,000 proxy statements, where our results can be accurately merged, based on URLs, with the data collected and shared by the UA library.[4] For those with discrepancies, we manually verify against the original sources to determine the correct values and then use these verified data points for comparison between the samples.

[3] In most of these cases, companies did not provide the CEO compensation in the pay disclosure section and instead referred readers to another section.

Panel B of Table 3 provides a comprehensive accuracy analysis by comparing collected pay ratios with those calculated from collected CEO pay and median employee pay figures. This analysis includes 9,749 cases where all three data elements were successfully collected. The findings indicate high consistency: in 9,567 cases (98.13%), the absolute difference between collected and calculated pay ratios is less than or equal to 1. Minimal discrepancies appear in the remaining cases: 34 cases (0.35%) have a difference between 1 and 2, 26 cases (0.27%) show a difference between 2 and 5, and 122 cases (1.25%) have a difference greater than 5. Differences under 2 are likely due to rounding, and the high percentage with differences most likely due to rounding validate both the model's extraction accuracy and the consistency of reported figures in proxy statements. Importantly, even absolute differences exceeding 5 do not necessarily indicate collection errors. Our investigation reveals that companies may apply aggressive rounding or occasionally miscalculate reported ratios.

Panel C examines 264 cases (2.68%), where the absolute difference is more than two (148 cases) or the difference is not available for evaluation because the LLM did not collect all three figures (116 cases). In many cases of the latter scenario, this is because not all three figures were disclosed in the source documents. We manually verify these 264 filings and report the discrepancy between the LLM-collected data and the company-disclosed data in Panel C of Table 3. The accuracy for CEO compensation, median employee pay, and pay ratio is 85.98%, 97.35%, and 96.59%, respectively, for these filings. Note that the greater discrepancy in CEO compensation is due to the fact that a significant number of firms do not provide total CEO compensation in the pay disclosure section but instead refer readers to the executive compensation table presented in another section. With these excluded, the accuracy for CEO compensation is comparable to those of median employee salary and pay ratio.

[4] The UA Library data (available at https://guides.lib.ua.edu/c.php?g=879087&p=9004058) does not provide URLs for all its observations, and matching based on company names and fiscal years can result in errors, weakening the comparison because discrepancies may be due to merging errors rather than differences in the actual data. It is also important to note that the data provided by the UA library appears to have rounded their compensation figures to whole dollars, and their pay ratios are not those provided in the actual disclosures but rather calculated based on the collected CEO pay and median employee pay. Therefore, we compare only the CEO compensation and median employee pay, and consider the data to be equal if the absolute difference is no more than one dollar.

Panel D compares the results of our LLM-collected data and those collected by the UA library against manually verified data, which serves as the ground truth. The results show that our LLM-collected data slightly outperforms the UA library's data in terms of accuracy. For CEO compensation, our accuracy is 99.68%, compared to the UA Library's 97.67%. Similarly, for median employee pay, our accuracy is 99.74%, while the UA Library's accuracy is 99.05%. We do not compare the accuracy for pay ratios, due to the limitations of the UA library data explained in footnote 4.

A conservative estimate of the overall accuracy based on CEO compensation is at least 99.27%, calculated as (9,567 cases from Panel B + (264×85.98%) cases from Panel C) / 9,865 total cases from Panel A. The accuracy is even higher for median employee pay and pay ratio. Moreover, all three metrics demonstrate an even higher level of accuracy when assessed based on the verified samples, as reported in Panel D.

Overall, these results demonstrate the LLM's reliability and effectiveness in automating pay ratio data collection from corporate filings. Only a small percentage of cases exhibit larger discrepancies or missing data, which may require additional verification or model refinements through further prompt engineering to handle varying report structures.

### 5.3 Results for CAMs

#### 5.3.1 Results of initial passage extraction

Panel A of Table 4 presents a summary of the initial Critical Audit Matters (CAM) extraction results. The results show that the regular expression (regex) approach is able to identify the beginning and end of audit reports in the vast majority of cases (96.84%). In some rare cases (3.16%) when only the heading of the CAM section is identified, we take a conservative approach by extracting 15,000 characters from the heading onwards. Overall, an average CAM section is 716 tokens long when successfully extracted from the audit report. If the end of the audit report is not identified, we extract 15,000 characters, which are approximately 2,134 tokens.

#### 5.3.2 Input tokens, and processing time and cost

Panel B of Table 4 provides a breakdown of the input tokens supplied to the LLM for collecting CAMs. The final prompt, which is provided in Fig. 2, consists of 836 tokens. A total of 12,499 CAM extracts were processed in batches of two extracts per request, resulting in 6,250 API requests.[5] The total input tokens, comprising both the prompt tokens (10.45 million) and the extract tokens (9.51 million), sum up to 19.96 million. The processing time, which includes error handling, is approximately 40 minutes. The total API cost amounts to approximately $8.

It is noteworthy that even though the total number of input tokens and number of requests are smaller compared to the pay ratio disclosures, the processing time for CAM collection is higher. This is because CAM collection requires re-generating the CAM, and an LLM typically processes input more quickly than generating output. Furthermore, the cost is also higher due to the fact that output tokens are significantly more expensive than input tokens (four times as high for our chosen model).

[5] We optimize processing efficiency by using a batch size of two, sending pairs of extracts within a single request along with the prompt. This approach reduces total processing costs by minimizing the number of times the prompt needs to be repeated. Unlike the task with pay ratio disclosures where cross-contamination between extracts could be problematic, our testing reveals no such issues for this specific task.

It is worth mentioning that CAM data is available through Audit Analytics at WRDS. However, the annual subscription fee can cost thousands of dollars, and to maintain access to the most up-to-date data, the subscription needs to be renewed regularly. This can be prohibitively expensive over the long run, making it difficult for researchers at financially constrained institutions to access this valuable resource. In contrast, our approach offers a highly cost-effective and time-efficient alternative. By leveraging an LLM, we are able to collect data from more than 12,000 annual reports, at a total cost of less than eight dollars. This exceptional efficiency demonstrates the potential of our method to democratize access to data for researchers who may not have the financial means to afford expensive subscriptions.

#### 5.3.3 Accuracy

We evaluate the accuracy of the GPT-collected and classified CAM data against a manually verified sample. First, our research assistant (RA), who is a master's student in a business program, manually collected CAM disclosures from a random sample of 500 10-K filings.[6] We then create a verified sample by comparing the GPT-collected data against the RA's manual collection. For cases where discrepancies exist between the GPT-collected and RA-collected data, the authors personally verify these instances to establish the ground truth. This two-stage verification process ensures a high-quality benchmark by identifying and correcting any potential errors in the initial manual collection. This approach allows us to not only evaluate the accuracy of our GPT-based methodology but also compare it to traditional manual data collection processes.

[6] Before the RA collected CAMs from the 500 samples for evaluation, we provided him with background information, detailed instructions, and training for the task. As a practice, he collected CAMs from a random sample of 100 proxy statements, and we compared his results with the LLM's results, providing feedback on the discrepancies to further improve his understanding of the task and ensure accuracy.

We employ cosine similarity to compare collected text against benchmarks. For clarity, similarity scores are rounded to the nearest 0.01, enabling clearer categorization while maintaining precision. However, for shorter texts like titles, minor character differences can disproportionately reduce similarity scores. This may result from encoding differences in special non-ASCII symbols, such as long dashes, between manually collected text saved in Excel format and LLM-collected data saved as plain text. Since each word carries greater weight in similarity calculations when word count is low, these encoding differences significantly impact scores. Consequently, lower similarity scores for shorter texts may reflect encoding differences rather than substantive content discrepancies.

We consider a match as perfect if the cosine similarity is one. As shown in Panel A of Table 5, out of 712 CAMs, 703 have a cosine similarity of one for the title, representing a 98.74% accuracy.[7] We see similarly outstanding results for "CAM descriptions" and "CAM procedures", at an accuracy of 98.74% and 97.75%, respectively.

Notably, most of the remaining cases have a cosine similarity of 0.99, often representing virtually identical text with only minor variations in spacing, punctuation, or formatting. When accounting for these near-perfect matches (cosine similarity ≥ 0.99) alongside perfect matches, the effective accuracy for all three metrics likely exceeds 99%. Even for non-perfect matches, the cosine similarity scores remain remarkably high, typically above 0.95, demonstrating that GPT model's output closely aligns with the verified sample. The model only failed to identify and collect information from two CAMs, representing just 0.28% of cases where titles, descriptions, and procedures were completely missed.

[7] There are 712 CAMs from the sample of 500 auditor reports because some reports contain multiple CAMs.

It is also worth mentioning that there are three instances of "zero" similarity scores for titles in the GPT-collected sample. These cases correspond to CAMs that originally had no titles. However, GPT demonstrated an additional capability by generating titles based on the descriptions of these CAMs, suggesting that GPT can be useful for in-depth analyses of CAM disclosure text, such as further classifying CAMs into categories.

Comparing GPT's performance to manual collection reveals comparable, and in some cases superior, results, as shown in Panel B of Table 5. GPT slightly outperforms manual collection in extracting titles and descriptions. However, manual collection shows a marginal advantage in procedure extraction due to GPT excluding in multiple cases the introductory sentences, which probably should be removed in later content analysis.[8] Interestingly, manual collection also missed two CAMs altogether, representing 0.28%, suggesting that both machine and human processes are susceptible to similar oversight errors. This parallel in error rate underscores that neither method is infallible, while also highlighting the comparable reliability of GPT-based extraction to traditional manual collection.

The accuracy analysis indicates that LLMs are not only a highly effective tool for CAM data collection but also show promise for more advanced applications in data analysis. Their performance matches or exceeds manual collection methods while offering significant efficiency gains and additional analytical capabilities. This is particularly important for researchers at disadvantaged institutions who may lack the funding to access expensive databases or hire research assistants for manual data collection. By providing an accurate and efficient alternative, LLMs can help level the playing field and enable a broader range of researchers to conduct meaningful analyses of qualitative textual information.

[8] An example of such introductory sentences is "The following are the primary procedures we performed to address this critical audit matter."

## 6. Discussion and conclusion

In this study, we explore the potential of democratizing access to costly datasets by leveraging recent advancements in GenAI. Using a capable and cost-effective LLM from OpenAI, we develop and evaluate an efficient approach for collecting large volumes of quantitative and qualitative data from unstructured text. Our approach proves highly efficient and cost-effective in that it can collect data from tens of thousands of documents in under an hour for less than US $10, with simpler tasks completed in minutes for just a few US dollars.

To promote research accessibility, we share our collected datasets of pay ratio and Critical Audit Matters (CAM) disclosures, both resulting from recent regulatory requirements. We provide detailed documentation of our methodology in Appendix C, enabling other researchers to replicate and adapt our approach. We hope this effort will contribute to the broader democratization of research by raising awareness and promoting the use of GenAI in ways that benefit disadvantaged researchers.

While our effort joins promising initiatives toward broader research democratization, several important challenges remain. Current LLMs are predominantly English-centric, limiting their effectiveness in analyzing non-English content (Filetti et al., 2024; Ghio, 2024), despite efforts to develop multilingual models that support both resource-rich and resource-limited languages (Chen et al., 2023). Additionally, market concentration, with OpenAI capturing 74.1 percent of the chatbot market through ChatGPT and Microsoft Copilot (Bailyn, 2024), poses challenges to truly democratic access. Furthermore, the cost of certain models remains prohibitively expensive, even for processing small amounts of data, and geographical restrictions prevent researchers in some countries from accessing certain LLMs.

Our findings also align with recent studies exploring the potential of LLMs to democratize various aspects of research and knowledge dissemination. For instance, Ni et al. (2023) introduce ChatReport, a tool that enhances LLMs with expert knowledge to automate the analysis of corporate sustainability reports, making this information more accessible and transparent. Similarly, Yue, Au, Au, and Iu (2023) demonstrate how ChatGPT can be used to explain complex financial concepts to non-financial professionals, empowering individuals to make informed investment decisions. Chang et al. (2023) provide empirical evidence of how democratized AI has transformed retail trading behavior. These studies, along with our own, highlight the potential of LLMs to bridge knowledge gaps and level the playing field in various domains.

However, as Ghio (2024) points out, the democratizing potential of LLMs is not without challenges, particularly in the context of language barriers and the dominance of English in research communication. Furthermore, as Ahmed and Wahed (2020) argue, the increasing computational intensity of modern AI research has led to a “compute divide,” where large firms and elite universities have an advantage due to their access to specialized equipment and resources. This divide threatens to “de-democratize” AI and presents an obstacle to truly inclusive knowledge production. Shashidhar, Chinta, Sahai, Wang, and Ji (2023) propose a solution to this problem by exploring cost-performance trade-offs in self-refined open-source models, demonstrating that even resource-constrained environments can leverage LLMs without compromising on performance or privacy.

Looking forward, we anticipate that increased market competition will foster more diverse and accessible research tools while driving down costs. As LLMs advance in multilingual capabilities and become more affordable, researchers worldwide may increasingly investigate broader geographical and cultural contexts. Despite present constraints, we remain optimistic about GenAI's potential to democratize research. We encourage policies that promote market competition, reduce access barriers, and support the development of more diverse and inclusive AI tools, particularly for researchers in underserved regions.

Finally, we call for research exploring how LLMs can enhance various aspects of the research process, from literature review and research design to data analysis and results interpretation. By automating routine tasks, researchers can dedicate more time to developing innovative ideas and theoretical insights, potentially accelerating scientific discovery and knowledge creation. As these technologies continue to evolve and become more sophisticated, we anticipate a transformative shift in research methodology that will enable a more diverse group of scholars to contribute meaningfully to their fields and address pressing societal challenges.

**Declaration of generative AI and AI-assisted technologies in the writing process:**

During the preparation of this work, the author(s) used ChatGPT and Claude to enhance language and readability. After using these tools, the author(s) reviewed and edited the content as needed and take(s) full responsibility for the content of this manuscript.

## References

Agrawal, A., & Goldfarb, A. (2008). Restructuring research: Communication costs and the democratization of university innovation. *American Economic Review*, *98*(4), Article 4. https://doi.org/10.1257/aer.98.4.1578

Ahmed, N., & Wahed, M. (2020). *The De-democratization of AI: Deep Learning and the Compute Divide in Artificial Intelligence Research* (No. arXiv:2010.15581). arXiv. https://doi.org/10.48550/arXiv.2010.15581

Anderson, C. (2004). The long tail. *Wired*. https://www.wired.com/2004/10/tail/

Andreoli-Versbach, P., & Mueller-Langer, F. (2014). Open access to data: An ideal professed but not practised. *Research Policy*, *43*(9), 1621–1633. https://doi.org/10.1016/j.respol.2014.04.008

Angrist, J., Azoulay, P., Ellison, G., Hill, R., & Lu, S. F. (2020). Inside job or deep impact? Extramural citations and the influence of economic scholarship. *Journal of Economic Literature*, *58*(1), Article 1. https://doi.org/10.1257/jel.20181508

Bailyn, E. (2024, August 26). Top generative AI chatbots by market share. *First Page Sage*. https://firstpagesage.com/reports/top-generative-ai-chatbots/

Bao, Y., & Datta, A. (2014). Simultaneously discovering and quantifying risk types from textual risk disclosures. *Management Science*, *60*(6), 1371–1391. https://doi.org/10.1287/mnsc.2014.1930

Beaudry, C., & Allaoui, S. (2012). Impact of public and private research funding on scientific production: The case of nanotechnology. *Research Policy*, *41*(9), Article 9. https://doi.org/10.1016/j.respol.2012.03.022

Bentley, J. W., Lambert, T. A., & Wang, E. (Ying). (2021). The effect of increased audit disclosure on managers' real operating decisions: Evidence from disclosing critical audit matters. *The Accounting Review*, *96*(1), 23–40. https://doi.org/10.2308/tar-2017-0486

Bergeaud, A., & Verluise, C. (2024). A new dataset to study a century of innovation in Europe and in the US. *Research Policy*, *53*(1), 104903. https://doi.org/10.1016/j.respol.2023.104903

Berninger, M., Kiesel, F., & Schnitzler, J. (2022). *Commercial data in financial research* (SSRN Scholarly Paper No. 3943132; Issue 3943132). https://doi.org/10.2139/ssrn.3943132

Beyer, B. D., Guragai, B., & Rapley, E. T. (2024). Critical audit matters: Recurring, nonrecurring, and intermittent. *Accounting Horizons*, 1–18. https://doi.org/10.2308/HORIZONS-2022-188

Bianchini, S., Müller, M., & Pelletier, P. (2022). Artificial intelligence in science: An emerging general method of invention. *Research Policy*, *51*(10), 104604. https://doi.org/10.1016/j.respol.2022.104604

Bland, C. J., & Ruffin, M. T. 4th. (1992). Characteristics of a productive research environment: Literature review. *Academic Medicine*, *67*(6), Article 6. https://journals.lww.com/academicmedicine/abstract/1992/06000/characteristics_of_a_productive_research.10.aspx

Boo, E., Low, K. Y., Shankar, P. G., & Tan, H.-T. (2024). Does discussing audit procedures in critical audit matter calibrate financial reporting risk assessments? *Accounting Horizons*, 1–13. https://doi.org/10.2308/HORIZONS-2022-040

Boone, A., Starkweather, A., & White, J. T. (2024). The saliency of the CEO pay ratio. *Review of Finance*, *28*(3), 1059–1104. https://doi.org/10.1093/rof/rfad039

Borgman, C. L. (2015). *Big data, little data, no data: Scholarship in the networked world*. MIT Press. http://ebookcentral.proquest.com/lib/oxford/detail.action?docID=3339930

Bourveau, T., Chowdhury, M., Le, A., & Rouen, E. (2025). *Human capital disclosures* (SSRN Scholarly Paper No. 4138543). https://papers.ssrn.com/abstract=4138543

Bozanic, Z., Dietrich, J. R., & Johnson, B. A. (2017). SEC comment letters and firm disclosure. *Journal of Accounting and Public Policy*, *36*(5), 337–357. https://doi.org/10.1016/j.jaccpubpol.2017.07.004

Brown, A. B., Wang, V. X., & Zhou, A. (2025). Employee perceptions of corporate culture and management forecast accuracy: Evidence from Glassdoor and ChatGPT. *Accounting Horizons*, 1–20. https://doi.org/10.2308/HORIZONS-2023-142

Burke, J. J., Hoitash, R., Hoitash, U., & Xiao, S. (Xia). (2023). The disclosure and consequences of U.S. critical audit matters. *The Accounting Review*, *98*(2), 59–95. https://doi.org/10.2308/TAR-2021-0013

Chang, A., Dong, X., Martin, X., & Zhou, C. (2023). *AI democratization, return predictability, and trading inequality* (SSRN Scholarly Paper No. 4543999). Social Science Research Network. https://doi.org/10.2139/ssrn.4543999

Chen, Z., Jiang, F., Chen, J., Wang, T., Yu, F., Chen, G., Zhang, H., Liang, J., Zhang, C., Zhang, Z., Li, J., Wan, X., Wang, B., & Li, H. (2023). *Phoenix: Democratizing ChatGPT across languages* (No. arXiv:2304.10453). arXiv. https://doi.org/10.48550/arXiv.2304.10453

Cheng, M., & Zhang, Y. (2023). Corporate stakeholders and CEO-worker pay gap: Evidence from CEO pay ratio disclosure. *Review of Accounting Studies*. https://doi.org/10.1007/s11142-023-09803-7

Cong, L. W., & Yang, S. (2024). *Understanding patenting disparities via causal human+machine learning* (SSRN Scholarly Paper No. 5171421). Social Science Research Network. https://doi.org/10.2139/ssrn.5171421

Cumming, D., Saurabh, K., Rani, N., & Upadhyay, P. (2024). Towards AI ethics-led sustainability frameworks and toolkits: Review and research agenda. *Journal of Sustainable Finance and Accounting*, *1*, 100003. https://doi.org/10.1016/j.josfa.2024.100003

Dai, R., Donohue, L., Drechsler, Q. (Freda), & Jiang, W. (2023). Dissemination, publication, and impact of finance research: When novelty meets conventionality. *Review of Finance*, *27*(1), Article 1. https://doi.org/10.1093/rof/rfac018

de Kok, T. (2023). *Generative LLMs and textual analysis in accounting: (Chat)GPT as research assistant?* (SSRN Scholarly Paper No. 4429658). https://doi.org/10.2139/ssrn.4429658

deHaan, E., Lawrence, A., & Litjens, R. (2024). Measuring investor attention using Google search. *Management Science*. https://doi.org/10.1287/mnsc.2022.02174

Demers, E., Wang, V. X., & Wu, K. (2024a). *Corporate human capital disclosures: Evidence from the first two years of the SEC's disclosure mandate* (SSRN Scholarly Paper No. 4153845). Social Science Research Network. https://doi.org/10.2139/ssrn.4153845

Demers, E., Wang, V. X., & Wu, K. (2024b). Measuring corporate human capital disclosures: Lexicon, data, code, and research opportunities. *Journal of Information Systems*, 1–24. https://doi.org/10.2308/ISYS-2023-023

Dong, M. M., Stratopoulos, T. C., & Wang, V. X. (2024). A scoping review of ChatGPT research in accounting and finance. *International Journal of Accounting Information Systems*, *55*, 100715. https://doi.org/10.1016/j.accinf.2024.100715

Dowling, M., & Lucey, B. (2023). ChatGPT for (finance) research: The Bananarama conjecture. *Finance Research Letters*, *53*, 103662. https://doi.org/10.1016/j.frl.2023.103662

Dundar, H., & Lewis, D. R. (1998). Determinants of research productivity in higher education. *Research in Higher Education*, *39*(6), Article 6. https://doi.org/10.1023/A:1018705823763

Dyer, T., Lang, M., & Stice-Lawrence, L. (2017). The evolution of 10-K textual disclosure: Evidence from Latent Dirichlet Allocation. *Journal of Accounting and Economics*, *64*(2), 221–245. https://doi.org/10.1016/j.jacceco.2017.07.002

El-Haj, M., Alves, P., Rayson, P., Walker, M., & Young, S. (2020). Retrieving, classifying and analysing narrative commentary in unstructured (glossy) annual reports published as PDF files. *Accounting and Business Research*, *50*(1), 6–34. https://doi.org/10.1080/00014788.2019.1609346

Filetti, S., Fenza, G., & Gallo, A. (2024). Research design and writing of scholarly articles: New artificial intelligence tools available for researchers. *Endocrine*, *85*(3), Article 3. https://doi.org/10.1007/s12020-024-03977-z

Ghio, A. (2024). Democratizing academic research with Artificial Intelligence: The misleading case of language. *Critical Perspectives on Accounting*, *98*, 102687. https://doi.org/10.1016/j.cpa.2023.102687

Giesecke, O. (2024). *AI at the frontier of economic research* (SSRN Scholarly Paper No. 4736003). https://doi.org/10.2139/ssrn.4736003

Goos, M., & Savona, M. (2024). The governance of artificial intelligence: Harnessing opportunities and mitigating challenges. *Research Policy*, *53*(3), 104928. https://doi.org/10.1016/j.respol.2023.104928

Hamermesh, D. S. (2018). Citations in economics: Measurement, uses, and impacts. *Journal of Economic Literature*, *56*(1), Article 1. https://doi.org/10.1257/jel.20161326

Heng, K., Hamid, Mo., & Khan, A. (2021). Factors influencing academics' research engagement and productivity: A developing countries perspective. *Issues in Educational Research*, *30*(3), Article 3. https://doi.org/10.3316/informit.465283943914964

Huang, X., Tan, L., Su, H., & Cheah, J. E.-T. (2025). Using deep learning conditional value-at-risk based utility function in cryptocurrency portfolio optimisation. *International Journal of Finance & Economics*, *n/a*(n/a). https://doi.org/10.1002/ijfe.70012

Jha, M., Qian, J., Weber, M., & Yang, B. (2024). *Harnessing generative AI for economic insights* (SSRN Scholarly Paper No. 4976759). Social Science Research Network. https://doi.org/10.2139/ssrn.4976759

Jha, M., Qian, J., Weber, M., & Yang, B. (2025). *ChatGPT and corporate policies* (SSRN Scholarly Paper No. 4521096). Social Science Research Network. https://doi.org/10.2139/ssrn.4521096

Karolyi, A. G. (2016). Home bias, an academic puzzle. *Review of Finance*, *20*(6), Article 6. https://doi.org/10.1093/rof/rfw007

Kim, E. H., Morse, A., & Zingales, L. (2006). What has mattered to economics since 1970. *Journal of Economic Perspectives*, *20*(4), Article 4. https://doi.org/10.1257/jep.20.4.189

Kim, E. H., Morse, A., & Zingales, L. (2009). Are elite universities losing their competitive edge? *Journal of Financial Economics*, *93*(3), Article 3. https://doi.org/10.1016/j.jfineco.2008.09.007

Kim, K., & Choi, S. B. (2017). Influences of creative personality and working environment on the research productivity of business school faculty. *Creativity Research Journal*, *29*(1), Article 1. https://doi.org/10.1080/10400419.2016.1239900

Klevak, J., Livnat, J., Pei, D. (Selina), & Suslava, K. (2023). Critical audit matters: Possible market misinterpretation. *AUDITING: A Journal of Practice & Theory*, *42*(3), 45–70. https://doi.org/10.2308/AJPT-2020-113

Korinek, A. (2023). Generative AI for economic research: Use cases and implications for economists. *Journal of Economic Literature*, *61*(4), 1281–1317. https://doi.org/10.1257/jel.20231736

Lertputtarak, S. (2008). *An investigation of factors related to research productivity in a public university in Thailand: A case study* [Other, Victoria University]. http://vuir.vu.edu.au/

Levich, S., & Knust, L. (2025). Discriminative meets generative: Automated information retrieval from unstructured corporate documents via (large) language models. *International Journal of Accounting Information Systems*, *56*, 100750. https://doi.org/10.1016/j.accinf.2025.100750

Lewis, P., Perez, E., Piktus, A., Petroni, F., Karpukhin, V., Goyal, N., Küttler, H., Lewis, M., Yih, W., Rocktäschel, T., Riedel, S., & Kiela, D. (2021). *Retrieval-augmented generation for knowledge-intensive NLP tasks* (No. arXiv:2005.11401). arXiv. https://doi.org/10.48550/arXiv.2005.11401

Li, F. (2010). The information content of forward-looking statements in corporate filings—A naïve Bayesian machine learning approach. *Journal of Accounting Research*, *48*(5), 1049–1102. https://doi.org/10.1111/j.1475-679X.2010.00382.x

Li, H., Gao, H. (Harry), Wu, C., & Vasarhelyi, M. A. (2024). Extracting financial data from unstructured sources: Leveraging large language models. *Journal of Information Systems*, 1–22. https://doi.org/10.2308/ISYS-2023-047

Muslu, V., Radhakrishnan, S., Subramanyam, K. R., & Lim, D. (2014). Forward-looking MD&A disclosures and the information environment. *Management Science*, *61*(5), 931–948. https://doi.org/10.1287/mnsc.2014.1921

Ni, J., Bingler, J., Colesanti Senni, C., Kraus, M., Gostlow, G., Schimanski, T., Stammbach, D., Vaghefi, S., Wang, Q., Webersinke, N., Wekhof, T., Yu, T., & Leippold, M. (2023). *chatReport: Democratizing sustainability disclosure analysis through LLM-based tools* (SSRN Scholarly Paper No. 4476733). https://doi.org/10.2139/ssrn.4476733

Przybyła, P., Brockmeier, A. J., Kontonatsios, G., Le Pogam, M.-A., McNaught, J., von Elm, E., Nolan, K., & Ananiadou, S. (2018). Prioritising references for systematic reviews with RobotAnalyst: A user study. *Research Synthesis Methods*, *9*(3), Article 3. https://doi.org/10.1002/jrsm.1311

Schwert, G. W. (2021). The remarkable growth in financial economics, 1974–2020. *Journal of Financial Economics*, *140*(3), Article 3. https://doi.org/10.1016/j.jfineco.2021.03.010

Shashidhar, S., Chinta, A., Sahai, V., Wang, Z., & Ji, H. (2023). *Democratizing LLMs: An exploration of cost-performance trade-offs in self-refined open-source models* (No. arXiv:2310.07611). arXiv. https://doi.org/10.48550/arXiv.2310.07611

Simisaye, A. O. (2019). A study of research productivity of the academic staff in research institutes in south-west Nigeria. *Samaru Journal of Information Studies*, *19*(2), Article 2. https://www.ajol.info/index.php/sjis/article/view/195467

Swanson, E. P. (2004). Publishing in the majors: A comparison of accounting, finance, management, and marketing. *Contemporary Accounting Research*, *21*(1), Article 1. https://doi.org/10.1506/RCKM-13FM-GK0E-3W50

Tan, W., Ren, Y., Chen, X. H., & Johan, S. (2025). *Tech-driven redress: The impact of Generative AI on consumer complaints in financial firms* (SSRN Scholarly Paper No. 5331357). Social Science Research Network. https://papers.ssrn.com/abstract=5331357

van Dalen, H. P., & Henkens, K. (2012). Intended and unintended consequences of a publish-or-perish culture: A worldwide survey. *Journal of the American Society for Information Science and Technology*, *63*(7), Article 7. https://doi.org/10.1002/asi.22636

Van Noorden, R., & Perkel, J. M. (2023). AI and science: What 1,600 researchers think. *Nature*, *621*(7980), Article 7980. https://doi.org/10.1038/d41586-023-02980-0

Wang, J., & Wang, V. X. (2025). *Assessing consistency and reproducibility in the outputs of large language models: Evidence across diverse finance and accounting tasks* (SSRN Scholarly Paper No. 5189069). Social Science Research Network. https://doi.org/10.2139/ssrn.5189069

Wanner, R. A., Lewis, L. S., & Gregorio, D. I. (1981). Research productivity in academia: A comparative study of the sciences, social sciences and humanities. *Sociology of Education*, *54*(4), Article 4. https://doi.org/10.2307/2112566

Yue, T., Au, D., Au, C. C., & Iu, K. Y. (2023). *Democratizing financial knowledge with ChatGPT by OpenAI: Unleashing the power of technology* (SSRN Scholarly Paper No. 4346152). Social Science Research Network. https://doi.org/10.2139/ssrn.4346152

## Fig. 1. Prompt for collecting pay ratio data.

You are an expert in CEO pay ratio disclosures required by the Dodd-Frank Act of 2010 in the USA.

For each excerpt from proxy statements of public companies, collect the following information:

1. Total CEO Compensation: As stated in the disclosure, used for calculating the pay ratio.

2. Total Median Employee Compensation: As stated in the disclosure, used for calculating the pay ratio.

3. CEO Pay Ratio: The ratio as reported in the disclosure.

**Special Instructions:**

- When provided with multiple numbered extracts (e.g., #1, #2, #3), treat each extract independently. Do not carry over information from one extract to another. Provide answers for each numbered extract based solely on the information contained within that specific extract.

- Collect compensation amounts and ratio as stated; do not perform calculations.

- If multiple amounts provided, extract those used for calculating the ratio.

- Return monetary amounts exactly as they are presented, including any specified units (e.g., use "75 thousand" or "35 million" if stated).

- Collect and return amounts in their original form, without assuming or adding thousands or millions unless explicitly mentioned.

- Do not round any figures (e.g., return "6.35" instead of "6".)

- Preserve the exact formatting of numbers, including all commas and decimal points. For example: return "20,399,972" as "20,399,972", not as "20399.972" or any other format; Return "86,933" as "86,933", not as "86.933" or any other format.

- Do not convert numbers to different representations (e.g., do not change to scientific notation or convert to thousands/millions).

- Do not add or remove zeros from the end of numbers.

- Return "Not Found" for any specific item (Total CEO Compensation, Total Median Employee Compensation, or Pay Ratio) that is not explicitly stated in the extract. Do not use a blank string, "Not Applicable", or any other placeholder - use only "Not Found" when the information is missing.

- Return "0" for total CEO compensation if the CEO is explicitly stated to receive no compensation or zero compensation.

- Return pay ratio as a single number (e.g., "20" for "20 to 1", "20:1", and "20 times").

- For percentage ratios, return the percentage value (e.g., "2.7%" instead of "2.7").

- Pay ratio may be zero or less than one in rare cases.

- For ratios like "43:13", return "43" as the pay ratio, ignoring the superscript note.

- Do not make up data for missing items.

- If multiple pay ratios or compensation figures are provided:

Extract all relevant ratios along with the corresponding CEO compensation and median employee compensation used to calculate each ratio;

Include each unique set of data (CEO compensation, median employee compensation, and corresponding pay ratio) as a separate entry;

Do not assume that the same total CEO compensation or median employee compensation applies to all ratios unless explicitly stated in the extract;

In the JSON output, include separate objects for each unique set of data within the list for that extract;

If the extract explicitly states that a particular compensation figure applies to multiple ratios, then you may use it accordingly.

- Focus on Relevant Sections: Pay special attention to sections with headings like "CEO Pay Ratio Disclosure", "Pay Ratio Disclosure", "Executive Compensation". If these headings are present, prioritize extracting information from the corresponding sections.

- If no clear "Pay Ratio Disclosure" section is found, search for the required information throughout the document, paying attention to paragraphs mentioning "median employee", "CEO compensation", and "pay ratio".

- Ignore Unrelated Content: If the text contains introductory or unrelated information, skip over it and concentrate on paragraphs likely to contain the required pay ratio details.

**Output Format:**

Return a single-line JSON object where:

Each key has the format "#N_X" where N is the extract number and X is the sequential number for multiple ratios

Each value is a three-element list: ["Total CEO compensation", "Total median employee compensation", "Pay ratio"]

Examples:

1. Single ratio in an extract:

{"#1_1": ["5,000,000", "50,000", "100"]}

2. Multiple ratios in an extract:

{"#1_1": ["5,000,000", "50,000", "100"], "#1_2": ["4,500,000", "45,000", "100"], "#1_3": ["5,000,000", "55,000", "91"]}

Do NOT return nested lists like this:

{"#1": [["5,000,000", "50,000", "100"], ["4,500,000", "45,000", "100"], ["5,000,000", "55,000", "91"]]}

3. Data from multiple extracts:

{"#1_1": ["5,000,000", "50,000", "100"], "#2_1": ["3,000,000", "60,000", "50"], "#3_1": ["4,000,000", "40,000", "100"], "#3_2": ["4,200,000", "42,000", "100"]}

[Placeholder for a list of source pay-ratio disclosure extracts]

## Fig. 2. Prompt for collecting CAM data.

```
You are an expert in Critical Audit Matters (CAMs) as required by the Public Company
Accounting Oversight Board (PCAOB) since 2018. For each CAM section extracted from an
auditor's report contained in 10-K filings, please gather the following information:

1. **Title**: The title of the CAM.

2. **Description**: The description of the CAM, providing context on why it is
critical.

3. **Audit Approach**: How the CAM was addressed during the audit.

**Special Instructions:**

- Exclude any introductory boilerplate paragraphs typically found in CAM sections.

- Each CAM is typically organized as follows:

  - "Title of CAM"

  - "Description Heading"

  - "Details of Description"

  - "Heading for How the CAM Was Addressed"

  - "Details for How the CAM Was Addressed"

- Use the headings to identify boundaries for each CAM and determine the corresponding
sentences or paragraphs for Title, Description, and Audit Approach.

- The Description section starts either with the "Description Heading" or immediately
after the Title if no heading is provided. It should include:

  - Background information about the transaction, event, or judgment involved,
sometimes referencing notes to the consolidated financial statements, though this is
not always mandatory.

  - Justifications or principal considerations leading the auditor to consider the
matter critical, including aspects of professional judgment regarding risk,
complexity, and potential for material misstatement.

  - The Description ends before the heading or details on how the CAM was addressed.

- Extract content between the boundaries exactly as it appears, removing any page
numbers (e.g., "F-2", "20"), table of contents entries (e.g., "Table of Contents"), or
footers/headers mixed in with paragraphs.

- Ensure that each CAM includes all three elements (Title, Description, Audit
Approach) unless the section is incomplete or truncated:

  - If a CAM lacks a title, proceed without it.

  - If the title contains phrases like "Refer to certain notes" (e.g., "Revenue -
Refer to Note 2 and Note 3 to the financial statements"), include this in the output.

  - If the title does not reference notes, do not add any, even if subsequent content
includes such references.

- Escape all double-quote characters (") in the output by adding a backslash (\).

- If any of the three elements are missing, return "Not Found" instead of leaving it
blank or using "Not Applicable."
```

- Capture each CAM separately, as reports may contain multiple CAMs.

- Ensure that for every CAM within an extract, all relevant content is classified under one of the categories (Title, Description, Audit Approach), with no content left uncategorized.

**Output Format:**

Return the data in the following JSON format, where each key is the extract ID (e.g., "#N_X") and the value is a list containing four elements:

1. The number of the CAM within the extract (e.g., "1", "2").

2. The title of the CAM.

3. The description of the CAM.

4. The audit approach for the CAM.

If an extract contains multiple CAMs, format the keys as "#N_1", "#N_2", etc., and ensure that each and every CAM is captured. Ensure the entire JSON object is output as a single line, with no extra spaces. Special characters such as double quotes and backslashes should be properly escaped.

**Example Output:**

{

    "#35_1": ["1", "Title of CAM 1", "Description of CAM 1", "Audit approach of CAM 1"],

    "#35_2": ["2", "Title of CAM 2", "Description of CAM 2", "Audit approach of CAM 2"],

    "#36_1": ["1", "Title of CAM 1", "Description of CAM 1", "Audit approach of CAM 1"]

}

**Final Special Instructions:**

- Do not generate or fabricate data for missing elements. If any element is not available, return "Not Found" instead.

[Placeholder for a list of source pay-ratio disclosure extracts]

## Table 1. Sample selection.

**Panel A: Pay ratio disclosures**

| | Number of Proxy Statements |
|---|---|
| All proxy statements filed with EDGAR over 2018-2023 | 33,425 |
| Proxy statements matched with Execucomp based on CIKs | 10,828 |
| Less: Proxy statements without pay ratio disclosures | (963) |
| Final sample of proxy statements with pay ratio disclosures | 9,865 |

**Panel B: Critical Audit Matters (CAMs)**

| | Number of 10-K Filings |
|---|---|
| All 10-Ks filed with EDGAR over 2019-2023 | 36,032 |
| Matched with Compustat and CRSP | 18,361 |
| 10-Ks without CAMs | (5,862) |
| Final sample of 10-Ks with CAMs | 12,499 |

**Table 2. Text extraction and LLM processing for pay ratio disclosure data.**

**Panel A: Initial text extracts from proxy statements for LLM processing**

| Extract Count | File Count | Percentage | Cumulative Percentage | Number of Extracts |
|---|---|---|---|---|
| 1 | 7,290 | 73.90 | 73.90 | 7,290 |
| 2 | 1,665 | 16.88 | 90.78 | 3,330 |
| 3 | 607 | 6.15 | 96.93 | 1,821 |
| 4 | 172 | 1.74 | 98.67 | 688 |
| 5 | 72 | 0.73 | 99.40 | 360 |
| 6 | 23 | 0.23 | 99.64 | 138 |
| 7 | 11 | 0.11 | 99.75 | 77 |
| 8 | 8 | 0.08 | 99.83 | 64 |
| 9 | 6 | 0.06 | 99.89 | 54 |
| 10 | 4 | 0.04 | 99.93 | 40 |
| 11 | 1 | 0.01 | 99.94 | 11 |
| 13 | 3 | 0.03 | 99.97 | 39 |
| 15 | 2 | 0.02 | 99.99 | 30 |
| 18 | 1 | 0.01 | 100.00 | 18 |
| Total | 9,865 | 100 | | 13,960 |

**Panel B: LLM task metrics (tokens, runtime, and cost)**

| Description | Value | Unit |
|---|---|---|
| Prompt tokens | 1,114 | tokens |
| Total extracts | 13,960 | extracts |
| Average tokens per extract | 1,821 | tokens/extract |
| Batch size | 1 | extracts/request |
| Number of requests | 13,960 | requests |
| Total prompt tokens | 15.55 | million tokens |
| Total extract tokens | 25.42 | million tokens |
| Total input tokens | 40.97 | million tokens |
| Total GPT processing time | 9 | minutes |
| Total API cost | $7 | USD |

Note: Number of requests = Total extracts / Batch size; Total prompt tokens = Prompt tokens * Number of requests; Total input tokens = Total prompt tokens + Total extract tokens. Processing time includes the time taken to handle API errors.

## Table 3. Results and accuracy of LLM-collected pay ratio data.

**Panel A: LLM collection summary**

| | Total Available | Collected | Collected% | Missing | Missing% |
|---|---|---|---|---|---|
| CEO Pay | 9,865 | 9,756 | 98.90% | 109 | 1.10% |
| Median Pay | 9,865 | 9,839 | 99.74% | 26 | 0.26% |
| Pay Ratio | 9,865 | 9,849 | 99.84% | 16 | 0.16% |

**Panel B: Internal consistency**

| Absolute Difference between Collected and Calculated Pay Ratios | Frequency | Percentage |
|---|---|---|
| <= 1 | 9,567 | 98.13% |
| 1-2 | 34 | 0.35% |
| 2-5 | 26 | 0.27% |
| >5 | 122 | 1.25% |
| Total | 9,749 | 100.00 |

**Panel C: Manual verification of cases missing data for internal consistency assessment**

| | CEO Pay | Median Employee Pay | Pay Ratio |
|---|---|---|---|
| Matched | 85.98% | 97.35% | 96.59% |
| Not Matched | 14.02% | 2.65% | 3.41% |
| Total documents | 264 | 264 | 264 |

**Panel D: Comparison with data manually collected by UA Library**

| Metric | Total Records | GPT Collected | GPT Accuracy | UA Library Collected | UA Library Accuracy |
|---|---|---|---|---|---|
| CEO Pay | 1,888 | 1,882 | 99.68% | 1,844 | 97.67% |
| Median Employee Pay | 1,903 | 1,898 | 99.74% | 1,885 | 99.05% |

## Table 4. Text extraction and LLM processing for CAM data.

**Panel A: Initial text extracts from 10-Ks for LLM processing**

| Category | Frequency | Percentage | Average Tokens | Total Tokens |
|---|---|---|---|---|
| CAM heading to end of auditor report | 12,104 | 96.84% | 716 | 8.67 M |
| CAM heading + 15,000 characters | 395 | 3.16% | 2,134 | 0.84 M |
| Total | 12,499 | 100.00% | 761 | 9.51 M |

Note: The "CAM heading to end of auditor report" category indicates that the text extract includes all characters from the CAM heading to the end of the auditor's report, while the "CAM heading + 15,000 characters" category indicates that the text extract consists of the first 15,000 characters following the CAM heading. "8.67 M" indicates 8.67 million.

**Panel B: LLM task metrics (tokens, runtime, and cost)**

| Description | Value | Unit |
|---|---|---|
| Prompt tokens | 836 | tokens |
| Total extracts | 12,499 | extracts |
| Average tokens per extract | 761 | tokens/extract |
| Batch size | 2 | extracts/request |
| Number of requests | 6,250 | requests |
| Total prompt tokens | 10.45 | million tokens |
| Total extract tokens | 9.51 | million tokens |
| Total input tokens | 19.96 | million tokens |
| Total GPT processing time | 40 | minutes |
| Total API cost | $8 | USD |

Note: Number of requests = Total extracts / Batch size; Total prompt tokens = Prompt tokens * Number of requests; Total input tokens = Total prompt tokens + Total extract tokens.

**Table 5. Results and accuracy of LLM-collected CAM Data.**

**Panel A: Comparison of GPT-collected data and verified data**

| Similarity | Title (N) | Desc (N) | Proc (N) | Title % | Desc (%) | Proc (%) |
|---|---|---|---|---|---|---|
| 1 | 703 | 703 | 696 | 98.74 | 98.74 | 97.75 |
| 0.99 | - | 2 | 7 | - | 0.28 | 0.98 |
| 0.98 | - | 2 | 2 | - | 0.28 | 0.28 |
| 0.97 | - | 2 | 2 | - | 0.28 | 0.28 |
| 0.96 | - | 1 | 1 | - | 0.14 | 0.14 |
| 0.95 | 1 | - | 1 | 0.14 | - | 0.14 |
| 0.86 | 1 | - | - | 0.14 | - | - |
| 0.72 | - | - | 1 | - | - | 0.14 |
| 0.62 | 1 | - | - | 0.14 | - | - |
| 0.46 | 1 | - | - | 0.14 | - | - |
| 0.00 | 3 | - | - | 0.42 | - | - |
| Missed | 2 | 2 | 2 | 0.28 | 0.28 | 0.28 |
| Total | 712 | 712 | 712 | 100.00 | 100 | 100 |

**Panel B: Comparison of RA-collected data and verified data**

| Similarity | Title (N) | Desc (N) | Proc (N) | Title % | Desc (%) | Proc (%) |
|---|---|---|---|---|---|---|
| 1 | 706 | 697 | 698 | 99.16 | 97.89 | 98.03 |
| 0.99 | - | 1 | 4 | - | 0.14 | 0.56 |
| 0.98 | - | 3 | 2 | - | 0.42 | 0.28 |
| 0.97 | - | 2 | 2 | - | 0.28 | 0.28 |
| 0.96 | - | 1 | 1 | - | 0.14 | 0.14 |
| 0.95 | - | 2 | - | - | 0.28 | 0.00 |
| 0.94 | - | 1 | 1 | - | 0.14 | 0.14 |
| 0.93 | - | 3 | 1 | - | 0.42 | 0.14 |
| 0.92 | 1 | - | - | 0.14 | - | - |
| 0.89 | - | - | 1 | - | - | 0.14 |
| 0.87 | 1 | - | - | 0.14 | - | - |
| 0.84 | 1 | - | - | 0.14 | - | - |
| 0.62 | 1 | - | - | 0.14 | - | - |
| Missed | 2 | 2 | 2 | 0.28 | 0.28 | 0.28 |
| Total | 712 | 712 | 712 | 100 | 100 | 100 |

Note: 'Similarity' represents the cosine similarity between GPT-collected (or RA-collected) and verified samples for CAM components (Title, Desc[ription], Proc[edure]). The verified sample serves as benchmark, constructed through RA collection with additional author verification. '(N)' shows item counts per similarity score. 'Missed' indicates CAMs unidentified by GPT or RA.

## Appendix A: Sample pay ratio disclosures

### Panel A: Free-form narrative

Source:
https://www.sec.gov/Archives/edgar/data/1159167/000115916722000019/a2022definitiveproxystatem.htm#i7b58200101764005979c7bfc4495e3ec_130

**2021 Pay Ratio**

Under the Dodd-Frank Wall Street Reform and Consumer Protection Act, the Company is required to disclose the annual total compensation of our median employee (excluding our chief executive officer), the annual total compensation of our principal executive officer, Chairman of the board of directors and chief executive officer, Colin Angle, and the ratio of these two amounts.

The Company selected January 1, 2022, the last day of our most recently-completed fiscal year, as the date upon which the median employee was identified. As of this date, the Company employed 1,372 employees globally, excluding 31 individuals that became employees as a result of the November 2021 acquisition of Aeris. The Company included all of our other full-time employees, part-time employees and interns, excluding the chief executive officer, in our analysis to identify the median employee. The Company did not elect to make any other exclusions as permitted under the SEC de minimis rule.

A Consistently Applied Compensation Measure was used to identify the median employee based on the sum of base pay/regular wages, overtime, bonus, commissions and equity grant date fair value. The Company elected to include bonus payments and equity awards given the broad participation rates in these programs across the employee population. Annualized salary rates for full-time employees and hourly pay rates and actual hours worked were used as reasonable estimates of salary/wages.

Using the compiled data, the Company determined that the 2021 annual total compensation of our median employee as of January 1, 2022 was $122,236 and Mr. Angle's annual total compensation for 2021 was $6,273,391, both of which were calculated in accordance with Item 402(c) of Regulation S-K. The ratio of these amounts was 51:1.

### Panel B: Bullet points + free-form narrative

Source: https://www.sec.gov/Archives/edgar/data/103145/000110465918018471/a18-2880_1def14a.htm

**Pay Ratio**

As required by Section 953(b) of the Dodd-Frank Wall Street Reform and Consumer Protection Act, and Item 402(u) of Regulation S-K, we are providing the following information about the relationship of the total annual compensation of our employees and the total annual compensation of Mr. Peeler, our Chief Executive Officer. The pay ratio included in this information is a reasonable estimate calculated in a manner consistent with Item 402(u) of Regulation S-K.

For 2017, our last completed fiscal year:

- the median of the total annual compensation of all employees (other than the CEO) was $108,356. For the purposes of calculating our CEO pay ratio, using the methodology described below, the total annual compensation of the median employee for 2017 was $141,390; and

- the total annual compensation of our CEO, as reported in the Summary Compensation Table above, was $2,402,882.

39

Table of Contents

Based on this information, for 2017 the ratio of the total annual compensation of Mr. Peeler, our CEO, to the median of the total annual compensation of all employees was 17.0 to 1.

The methodology and the material assumptions, adjustments, and estimates that we used to identify the median of the total annual compensation of all our employees, as well as to determine the total annual compensation of the "median employee," were as follows:

1. We determined that as of October 1, 2017, our employee population consisted of approximately 1,031 individuals working at Veeco and its subsidiaries, with 70% of these individuals located in the United States.

## Appendix A: Sample pay ratio disclosures (Continued)

### Panel C: Tabulated form

Source: https://www.sec.gov/Archives/edgar/data/884219/000156459019010690/vvi-def14a_20190516.htm#CEO_PAY_RATIO

**CEO PAY RATIO**

As required by Section 953(b) of the Dodd-Frank Wall Street Reform and Consumer Protection Act ("Dodd-Frank"), we are providing the following disclosure that compares the annual total compensation of our "median employee" to the annual total compensation of our Principal Executive Officer, our CEO.

To determine our median employee, we included base salary, which is paid in the form of hourly wages, and commissions paid, as the consistently applied compensation measure for all employees. We selected these pay elements because they were the most consistently paid across our organization. We reviewed compensation as of October 13, 2017, to determine our median employee. As of that date, we had 18,223 employees according to the definition provided under Dodd-Frank, though not all of these employees were actively working at that time. In determining our median employee, we excluded employees from countries that represent 5% or less of our global headcount. The excluded countries were Hong Kong, The Netherlands, Romania, Switzerland, and the United Arab Emirates. Combined, this population represented 192 employees, or less than 1% of our total headcount.

We applied statistical sampling to develop a narrow range of employees around our estimated median pay for 2017 of $4,662. Using a range of pay within 10% of this estimated median, we identified 591 employees. We then

VIAD Viad Corp | EXECUTIVE COMPENSATION 47

conducted further analysis of prior years' earnings to identify 30 employees from this group with relatively stable earnings over the past several years. Finally, from this list of 30 employees, we selected our median employee, who was a part-time employee in 2017, and also in 2018, and who was represented by a union.

Having determined our median employee, we collected additional elements of pay pursuant to Item 402(c)(2)(x) of Regulation S-K, which is the same methodology used to determine compensation for our CEO and our other NEOs in the Summary Compensation Table in this Proxy Statement. We reported the results of our analysis in our proxy statement filed on April 4, 2018.

For our 2018 fiscal year, Dodd-Frank allows us to review compensation for the same median employee we identified in 2017. We verified that the employee was employed generally on the same basis as in 2017. We also confirmed that there was no compelling reason to select a different median employee for our 2018 disclosure. Accordingly, the table below provides the annual total compensation for our median employee and for our CEO for 2018, as well as the ratio of our CEO's total compensation to that of the median employee.

| **CEO Pay Ratio – includes part-time and seasonal employees:** | |
|---|---|
| 2018 Total Annual Compensation – Median Employee | $5,501 |
| 2018 Total Annual Compensation – Steven W. Moster, CEO | $3,741,915 |
| Ratio of CEO Compensation to the Median Employee | 680:1 |

### Panel D: Tabulated form + free-form narrative

Source: https://www.sec.gov/Archives/edgar/data/1095073/000109507322000007/proxy2022.htm

**CEO PAY RATIO DISCLOSURE**

| Fiscal Year | | 2021 | | 2021 |
|---|---|---|---|---|
| **Employee** | | Median Employee | | CEO |
| **Annual Base Salary** | $ | 133,100 | $ | 1,250,000 |
| **Bonus Paid March 2022** | $ | 14,000 | $ | 3,000,000 |
| **Res Share Value Granted Feb. 2021** | $ | 0 | $ | 2,000,000 |
| **Perf Share Target Value Granted Feb. 2021** | $ | 0 | $ | 2,000,000 |
| **Pension Value and Nonqualified Deferred Comp Earnings PY 2021** | $ | 0 | $ | 0 |
| **All Other Compensation PY 2021** | $ | 4,176 | $ | 614,322 |
| **Total Comp** | $ | 151,276 | $ | 8,864,322 |

In 2021, the ratio of the total annual compensation of our CEO to the median compensation of our employees was 58.60 to one.

**Methodology**

- Date selected to determine employee population for purposes of identifying the median employee– December 1, 2021.
- Median employee identified using Total Compensation, which includes base salary, bonus, and stock awards (if any) as well as any other compensation.

# Appendix B: Sample Critical Audit Matter (CAM) disclosures

## Panel A: Free-form narrative

Source: https://www.sec.gov/ix?doc=/Archives/edgar/data/896156/000143774921020534/eth20210806_10k.htm

***Critical Audit Matter***

The critical audit matter communicated below is a matter arising from the current period audit of the consolidated financial statements that was communicated or required to be communicated to the audit committee and that: (1) relates to accounts or disclosures that are material to the consolidated financial statements and (2) involved our especially challenging, subjective, or complex judgments. The communication of a critical audit matter does not alter in any way our opinion on the consolidated financial statements, taken as a whole, and we are not, by communicating the critical audit matter below, providing a separate opinion on the critical audit matter or on the accounts or disclosures to which it relates.

*Assessment of the carrying value of retail design center long-lived assets*

As discussed in Note 3 to the consolidated financial statements, the Company reviews long-lived assets for impairment whenever events or changes in circumstances indicate that the carrying value of these assets may not be recoverable. If the sum of the estimated undiscounted future cash flows over the remaining life of the primary asset is less than the carrying value, the Company recognizes a loss equal to the difference between the carrying value and the fair value. As of June 30, 2021, property, plant and equipment, net, was $231.4 million and the Company recognized impairment charges of $0.6 million for the fiscal year ended June 30, 2021.

We identified the assessment of the carrying value of retail design center long-lived assets as a critical audit matter. Specifically, complex auditor judgment was required to assess the sales growth rates used to estimate the forecasted cash flows as they involved a high degree of subjectivity.

The following are the primary procedures we performed to address this critical audit matter. We evaluated the design and tested the operating effectiveness of certain internal controls over the Company's retail design center impairment assessment process, including controls related to the development of the sales growth rates. We evaluated the Company's sales growth rates by (1) comparing them to historical results, the Company's future operating plans, existing retail orders backlog, and industry reports, and (2) performing sensitivity analyses.

## Panel B: Structured format with component headers

Source: https://www.sec.gov/ix?doc=/Archives/edgar/data/1568100/000156810022000014/pd-20220131.htm

***Revenue Recognition***

*Description of the Matter*

The Company's revenue totaled $281.4 million for the year ended January 31, 2022. As described in Note 2 to the consolidated financial statements, the Company primarily generates revenue from cloud-hosted subscription fees, with the majority of its revenue recognized from such arrangements. In order to recognize revenue, the Company evaluates whether promises made to customers represent distinct performance obligations, the appropriate measure of the transfer of control and when the transfer of control has occurred. These assessments can require significant judgment, particularly when contracts include non-standard terms.

Auditing the Company's accounting for revenue recognition was complex because certain of the Company's revenue agreements contained non-standard contractual terms that required significant auditor judgement to determine if distinct performance obligations were created. The proper identification of performance obligations in the Company's revenue arrangements could have a significant impact on the timing of revenue recognition and the disclosures.

*How We Addressed the Matter in Our Audit*

We obtained an understanding, evaluated the design, and tested the operating effectiveness of controls over the Company's process to identify and evaluate performance obligations including identification and consideration of non-standard contractual terms, the transaction price, and the measure of progress of the transfer of control.

Our audit procedures included, among others, reading a sample of contracts and evaluating whether management appropriately identified and considered terms within those documents that would affect revenue recognition, and testing the Company's evaluation of standalone selling price for its performance obligations. We also evaluated the completeness and accuracy of the underlying data used in management's determination of standalone selling price and the recorded deferred revenue and revenue amounts.

## Appendix C: Detailed implementation of methodology

### C1 Technical challenges

Both CEO pay ratio data and Critical Audit Matters (CAMs) data present significant technical challenges for automated collection. The CEO pay ratio compares a company's CEO total annual compensation to the median annual compensation of all other employees. While companies must disclose three standardized elements in their proxy statements (i.e., CEO total compensation, median employee compensation, and their ratio), the presentation formats vary considerably across firms.

As illustrated in Appendix A, companies employ diverse approaches ranging from narrative descriptions to structured tables and hybrid formats. This heterogeneity in disclosure formats complicates the development of universal extraction algorithms. Traditional rule-based systems and supervised machine learning models struggle with these variations, requiring substantial manual annotation while achieving limited accuracy and generalizability.

Critical Audit Matters (CAMs) extraction faces similar challenges. Although CAMs follow a general structure, organized as title, issue description, criticality rationale, and audit procedures, their presentation varies markedly across companies. Some reports use continuous narrative formats, while others employ tabular structures with headings such as "Description of the Matter" and "How We Addressed the Matter in the Audit" (see Appendix B). Moreover, when reports contain multiple CAMs, the extraction process must be able to accurately identify and decompose each matter into its constituent elements.

These complexities necessitate advanced natural language processing techniques, particularly state-of-the-art large language models (LLMs), which can adapt to diverse formatting patterns and extract structured data from heterogeneous disclosure formats without extensive manual preprocessing.

### C2 Advantages of LLMs

Large Language Models (LLMs) offer a promising solution to the challenges of extracting data from heterogeneous corporate disclosures and other documents. Pre-trained on vast document corpora and fine-tuned through instruction-based learning, these models excel at recognizing patterns across diverse presentation styles while maintaining contextual understanding. This capability enables them to extract CEO pay ratio components from varied formats and accurately identify and decompose Critical Audit Matters (CAMs) into their constituent elements, including title, description, criticality rationale, and audit procedures.

The key advantage of LLMs lies in their adaptability. Unlike traditional extraction methods, LLMs can distinguish core data from supplementary context and process hybrid presentation formats without extensive preprocessing. Whether analyzing narrative descriptions, structured tables, or mixed formats in pay ratio disclosures, or handling single versus multiple CAMs in auditor reports, these models maintain consistent performance. In addition, LLMs provide unprecedented scalability. They can process tens of thousands of financial documents in hours rather than months, extracting structured data that would otherwise require extensive manual collection.

## C3 Small-scale experiments with ChatGPT

We first conducted small-scale experiments using ChatGPT's user interface to evaluate its effectiveness in extracting CEO pay ratio disclosures and CAM disclosures. Using zero-shot learning, we provided ChatGPT with concise prompts to extract compensation data and CAM components. The experiments yielded impressive results: ChatGPT successfully extracted pay ratios across varied formats (Fig. A-1) and accurately decomposed CAMs into their title, description, and procedure sections (Fig. A-2).

Notably, the model maintained high performance even when document formatting was lost during copy-paste, demonstrating its robustness in handling unstructured data. These encouraging results, which showed ChatGPT's ability to navigate complex financial disclosures without additional training or examples, led us to scale up the experiments using the API for processing large volumes of documents.

## C4 Identifying relevant sections in large documents

The successful implementation of LLMs for extracting pay ratio data and CAM data from corporate documents requires addressing a fundamental challenge: locating and extracting the relevant text from source documents before feeding it to the language model. Proxy statements and 10-K filings often contain extensive amounts of text, with the sections relevant to our task buried within unrelated content. To address this challenge efficiently while minimizing costs, we developed an approach based on careful examination of document structures.

Our analysis revealed that sections containing pay ratio disclosures and CAMs typically follow predictable patterns within these corporate documents. Pay ratio disclosures in proxy statements generally appear under dedicated headings or can be identified through mentions of median employee salary and related terminology. Similarly, CAMs are contained within auditor reports that follow specific structures with relatively standardized language. This relative consistency in document structure enables us to employ regular expressions for precise location and extraction of relevant text sections. Regular expressions offer several advantages for this task: they allow precise targeting of specific sections, preserve the integrity of disclosures by extracting complete sections rather than fragments, and provide computational efficiency that facilitates scaling to large document collections.

We considered an alternative approach commonly used for processing large documents: chunking combined with embedding and retrieval models. This technique involves breaking down documents into smaller, manageable segments—a necessary step given that proxy statements average nearly 40,000 words and 10-K filings exceed 65,000 words. These segments are then transformed into dense vector embeddings and indexed in a vector database for similarity-based retrieval. However, this approach presents several drawbacks. The computational overhead of processing and indexing large text volumes is substantial, and the chunking process risks fragmenting information across chunk boundaries. Additionally, sophisticated scoring mechanisms are required to ensure accurate retrieval of relevant content, adding system complexity.

## C5 Framework and procedures

Building on the methodological considerations described above, we developed a comprehensive framework for extracting CEO pay ratios and Critical Audit Matters using LLMs. Fig. A-3 illustrates our ten-step process. We designed Python scripts to automate all steps except prompt engineering, which required manual iteration and refinement.

(1) Download crawler index URLs: This step involves downloading the index files that contain crawler URLs for proxy statements and 10-K filings.

(2) Extract URLs for filings in HTML format: This step navigates to the webpages corresponding to the crawler URLs and extract the URLs for the HTML versions of the filings.

(3) Download HTML filings: Using the extracted URLs, we download the filings in HTML format.

(4) Parse filings: We process the downloaded HTML files to extract the text content by removing HTML tags, scripts, and formatting elements.

(5) Develop regex for CEO pay ratio and CAM: We create regular expressions to precisely locate and extract the sections related to CEO pay ratios from proxy statements and CAMs from 10-K filings.

(6) Extract sections using regex: By applying the developed regular expressions to the parsed filing text, we isolate and extract the specific sections containing the CEO pay ratio and CAM content.

(7) Perform prompt engineering: We craft prompts to guide the LLM in identifying and extracting target data from the text sections. Through iterative testing on sample extracts, we refine these prompts to maximize extraction accuracy and consistency.

(8) Submit all extracts with final prompts to OpenAI API: We send the extracted sections along with the final prompts to the OpenAI API. The API processes the text using the specified language model and returns the collected data in the specified format.

(9) Parse, clean, and merge the data from API responses: We parse the API responses to obtain the relevant data and consolidate the results into a structured format.

(10) Evaluate the accuracy of results: Finally, we assess the quality and accuracy of the extracted CEO pay ratio and CAM data to ensure the reliability of the process.

In the next few sections, we describe the key steps in greater detail.

## C6 Extracting pay ratio and CAM sections from source documents

The extraction process employs a two-stage approach to locate pay ratio disclosures. First, it identifies specific pay ratio headings and extracts an asymmetric window of text: 1,000 characters preceding and 7,000 characters following each heading. If no relevant heading is found, the process searches for mentions of “median employee” and applies the same extraction window. This asymmetric approach reflects our observation that relevant information typically follows rather than precedes these reference points. In cases where multiple potential disclosure sections are found, we extract all of them to maximize the likelihood of capturing the required information.

This approach allows for redundancy, which is necessary given the varied presentation styles of pay ratio disclosures.

Extracting CAM sections is relatively less challenging due to the more consistent formatting of auditor reports across firms. The audit report typically begins with "We have audited the accompanying consolidated financial statements" and concludes with "We have served as the auditor of the company since [year]." After identifying these boundaries, we extract the entire audit report and subsequently isolate the CAM section based on the "Critical Audit Matter" heading. In cases where a report's end is not easily identifiable, we extract 15,000 characters (equivalent to more than 2,000 tokens) following the "Critical Audit Matter" heading to ensure capturing the full CAM section, particularly when multiple matters are present.

## C7 Model choice

In this study, we choose the GPT-4o-mini model (specifically, gpt-4o-mini-2024-07-18) as the foundation for our experiments, leveraging its optimal balance of performance and cost-effectiveness.[9] This model, introduced as a more efficient alternative to GPT-3.5-Turbo, offers enhanced capabilities at a lower cost, making it particularly suitable for our research objectives.

The model's features align well with our study's requirements. Its 128K context window enables analysis of longer text sequences and complex contextual relationships, crucial for our research methodology. This large context window allows us to submit multiple pay ratio or CAM extracts in a single request, resulting in significant cost savings on prompt tokens.

The model has a maximum output capacity of 16,384 tokens. This large output limit provides confidence that the model will extract full CAMs, which can be lengthy, especially when multiple CAM extracts are provided. Without this capacity, the model might stop generating prematurely, resulting in incomplete results. In addition, the model's pricing structure ($0.15 per 1M input tokens and $0.60 per 1M output tokens) allows for cost-effective implementation.

## C8 Prompt engineering

### C8.1 Prompt for pay ratio data collection

Prompt engineering is a technique that involves crafting prompts in a way that elicits the desired result from an LLM. As LLMs become more advanced, there is a growing belief that prompt engineering has become less important. However, our experiment suggests that prompt engineering remains a crucial aspect of working with LLMs.

Our approach to prompt engineering begins with providing a relatively simple prompt and experimenting with a small dataset to observe the results and identify cases where the LLM fails to produce the expected output. This process is iterative, requiring multiple rounds of refinement to finalize a prompt that consistently yields accurate results. Through this iterative process, we develop the prompt for our large-scale experiments in collecting pay-ratio related data.

---

[9] It is noteworthy that our initial experiments with ChatGPT may have utilized GPT-4o, as OpenAI provides limited free access to GPT-4o, even for non-paying users. This could account for the subsequent decrease in performance when we transitioned to the API using a more cost-effective model with a basic prompt. However, we were able to enhance performance through the use of a refined prompt.

As shown in Fig. 1, our final prompt instructs the model to extract three data points from CEO pay ratio disclosures: total CEO compensation, median employee compensation, and the pay ratio. The design handles various formatting scenarios (e.g., "30 thousand," "20:1," "2.7%") and special cases, returning "Not Found" for missing information to prevent data fabrication. Results are delivered in JSON format for efficient parsing subsequently.

To optimize costs, we batch multiple text extracts in single API calls rather than processing them individually, sharing instructions across extracts to minimize input tokens. We also eliminate unnecessary whitespace in outputs since output tokens cost more than input tokens do. The prompt uses structured markdown with headers, numbered lists, and code formatting to guide the model through complex extraction scenarios while maintaining consistency. This approach balances accuracy, efficiency, and cost-effectiveness in large-scale data collection.

### C8.2 Prompt for CAM collection

To instruct the model for extracting CAMs and classifying them into components, we initially provide a simple prompt along the lines of "extracting the title, description, and procedure for each critical audit matter (CAM) from the text provided." However, our chosen model does not fully extract the required data with this basic prompt.

To improve the model's performance, we gradually enhance the prompt by providing more detailed instructions. During this process, we find that utilizing the ChatGPT user interface is particularly helpful in optimizing the prompt by asking it refine the prompt for an LLM. By iteratively refining the instructions and testing the model's output, we create the final prompt, as shown in Fig. 2.

Our final prompt begins with background on Critical Audit Matters (CAMs) and PCAOB requirements, and then provides specific instructions for extracting three elements from each CAM: title, description, and audit approach. The design addresses various scenarios including missing elements, formatting issues, and multiple CAMs within single extracts. Output is specified in JSON format with examples provided to ensure consistency.

The detailed instructions emerged from iterative testing on sample CAM sections, identifying where the model initially struggled. We provide guidance similar to training a research assistant, i.e., describing each element's content and how to identify boundaries within CAM sections. The prompt explicitly prevents data fabrication by instructing the model to indicate when elements are missing rather than generating content.

We use structured markdown formatting to enhance clarity: headers separate major sections, bullet points list specific requirements, bold text highlights key terms, and code blocks present output examples. This approach, combining detailed instructions with clear formatting, effectively adapts the model to accurately extract and classify CAM content while maintaining data integrity.

## C9 Parallel API processing

To efficiently process large volumes of text using the OpenAI API, we utilize the parallel processing code provided by OpenAI.[10] This code offers several inherent features that make it well-suited for our purpose, including streaming requests from files to handle large datasets, making concurrent requests to maximize throughput, throttling requests to stay within rate limits, retrying failed requests to ensure data completeness, and logging errors for effective troubleshooting.

[10] https://github.com/openai/openai-cookbook/blob/main/examples/api_request_parallel_processor.py

Because the original OpenAI code was designed for generating text embeddings, we significantly modified it to suit our specific data collection tasks. Unlike text embedding generation, which is relatively straightforward, data collection requires robust error handling and management of unexpected behaviors. Key modifications include:

- Data structure: We transitioned from JSONL to CSV files and DataFrames, facilitating easier debugging, troubleshooting, and prompt engineering. This change allows for convenient data inspection in Excel, particularly valuable during the prompt engineering phase.
- Input handling: Instead of reading requests from a file, our modified code accepts a list of dynamically constructed prompts with input data.
- API endpoint: We now utilize the chat completions endpoint (`https://api.openai.com/v1/chat/completions`) to interact with the language model for generating responses based on chat conversations.[2]
- Duplicate prevention: We introduced a mechanism to avoid duplicate processing by tracking completed inputs based on document IDs in the output file. This enhancement allows for efficient resumption of processing in case of interruptions, particularly useful when dealing with response formatting errors or unexpected terminations.
- Error logging: We implemented separate logging for API errors (such as response timeouts or over-capacity issues) and response format errors (e.g., incorrect JSON format in the API response). This separation enables more targeted troubleshooting and error analysis.

These modifications collectively transform the original script into a robust, efficient tool tailored for our specific data collection and processing needs, while preserving the core functionality of asynchronous API calls, request throttling, and failure retry mechanisms.

When configuring the OpenAI API for our data collection tasks, we set the model's temperature to zero. This ensures reproducibility and consistency in the generated output. Unlike creative writing or other text generation tasks that benefit from diversity and creativity, data collection requires precise and deterministic results. By setting the temperature to its lowest possible value, we minimize the randomness in the model's output, making it more suitable for our specific task.

In addition to the temperature, we set a seed value for the model. Although OpenAI does not guarantee fully deterministic results, setting a seed helps the model do its best to produce consistent output across multiple runs. By using the same seed value, we can expect the model to generate highly consistent results each time, provided that the input data and prompts remain unchanged.

It is worth sharing that our experiments uncovered several challenges that required effective mitigation strategies. First, when processing pay ratio disclosures, multiple extracts in one prompt occasionally resulted in cross-contamination, where data from one extract was incorrectly attributed to another. We resolved this by reducing the batch size to one extract per query. While this approach slightly increased processing costs due to prompt repetition, the improved accuracy justified the additional expense. For CAM collection, cross-contamination proved not an issue, likely because the task involves simpler categorization and content regeneration.

Second, longer extracts sometimes reduced collection accuracy for pay ratio data, likely because relevant information was embedded within extensive unrelated content. We developed two approaches to address this challenge: (1) Conservative truncation. For unsuccessful initial attempts, we removed 1,000 characters from each end of the 8,000-character extracts. This moderate approach preserved 6,000 characters, maintaining sufficient context for accurate data extraction while eliminating potentially distracting peripheral content. (2) Progressive truncation: For more persistent cases not resolved by single truncation, we implemented a gradual truncation strategy that began with more aggressive truncation and progressively reduced it as needed. This flexible approach allowed us to optimize the balance between maintaining adequate context and improving extraction accuracy. While we initially maintained longer extracts (8,000 characters) during data preparation to ensure complete coverage, these truncation strategies proved essential for handling cases where excessive context hindered rather than helped the extraction process.

Third, for challenging extracts, we improved model performance by including a sample pay ratio disclosure, which helped clarify task requirements and output format. We implemented detection mechanisms to prevent cross-contamination, ensuring the model's output corresponded to the target extract rather than the example.

While hallucination was not found to be an issue in our experiments, we identified straightforward verification mechanisms that could be implemented if needed. For numerical data collection tasks, hallucination can be readily detected through a two-step verification: first confirming that the extracted numbers exist in the source text, then verifying their contextual relevance. For instance, median pay figures should appear near terms like "median employee" or "workers," while pay ratios should be proximate to the word "ratio." This simple two-step verification provides a robust safeguard against potential hallucination in numerical extraction tasks. For text extraction tasks such as CAM collection, hallucination risk is inherently lower since the model simply identifies and reproduces existing content verbatim, although similar verification methods could be applied if necessary.

## Fig. A-1 Extracting CEO pay ratios using ChatGPT.

Prompt: What are the CEO compensation, median employee compensation, and pay ratio based on the disclosure provided? Return a JSON object.

### Panel A: Free-from narrative

What are the CEO compensation, median employee compensation, and pay ratio based on the disclosure provided? Return a JSON object. Using the compiled data, the Company determined that the 2021 annual total compensation of our median employee as of January 1, 2022 was $122,236 and Mr. Angle's annual total compensation for 2021 was $6,273,391, both of which were calculated in accordance with Item 402(c) of Regulation S-K. The ratio of these amounts was 51:1.

json

```json
{
  "CEO_compensation": 6273391,
  "median_employee_compensation": 122236,
  "pay_ratio": "51:1"
}
```

### Panel B: Bullet points + free-form narrative

For 2017, our last completed fiscal year:

- the median of the total annual compensation of all employees (other than the CEO) was $108,356. For the purposes of calculating our CEO pay ratio, using the methodology described below, the total annual compensation of the median employee for 2017 was $141,390; and
- the total annual compensation of our CEO, as reported in the Summary Compensation Table above, was $2,402,882.

Table of Contents

Based on this information, for 2017 the ratio of the total annual compensation of Mr. Peeler, our CEO, to the median of the total annual compensation of all employees was 17.0 to 1.

json

```json
{
  "CEO_compensation": 2402882,
  "median_employee_compensation": 141390,
  "pay_ratio": "17:1"
}
```

### Panel C: Tabulated form

For our 2018 fiscal year, Dodd-Frank allows us to review compensation for the same median employee we identified in 2017. We verified that the employee was employed generally on the same basis as in 2017. We also confirmed that there was no compelling reason to select a different median employee for our 2018 disclosure. Accordingly, the table below provides the annual total compensation for our median employee and for our CEO for 2018, as well as the ratio of our CEO's total compensation to that of the median employee.

CEO Pay Ratio – includes part-time and seasonal employees:

| | |
|---|---|
| 2018 Total Annual Compensation – Median Employee | $5,501 |
| 2018 Total Annual Compensation – Steven W. Moster, CEO | $3,741,915 |
| Ratio of CEO Compensation to the Median Employee | 680:1 |

json

```json
{
  "CEO_compensation": 3741915,
  "median_employee_compensation": 5501,
  "pay_ratio": "680:1"
}
```

## Fig. A-1 Extracting CEO pay ratios using ChatGPT (Continued)

**Panel D: Tabulated form + free-form narrative**

| | |
|---|---|
| CEO PAY RATIO DISCLOSURE<br>________________________________________________<br>Fiscal Year 2021 2021<br>Employee Median Employee CEO<br>Annual Base Salary $ 133,100 $ 1,250,000<br>Bonus Paid<br>March 2022 $ 14,000 $ 3,000,000<br>Res Share Value Granted<br>Feb. 2021 $ 0 $ 2,000,000<br>Perf Share Target Value Granted<br>Feb. 2021 $ 0 $ 2,000,000<br>Pension Value and Nonqualified Deferred Comp Earnings<br>PY 2021 $ 0 $ 0<br>All Other Compensation<br>PY 2021 $ 4,176 $ 614,322<br>Total Comp $ 151,276 $ 8,864,322<br><br>In 2021, the ratio of the total annual compensation of our CEO to the median compensation of our employees was 58.60 to one. | json<br>{<br>"CEO_compensation": 8864322,<br>"median_employee_compensation": 151276,<br>"pay_ratio": "58.6:1"<br>} |

## Fig. A-2 Extracting CAMs using ChatGPT

Prompt: Extract the critical audit matter and break it down into the title, description, and procedure

### Panel A: Free-form narrative

Here is the extracted critical audit matter broken down into the title, description, and procedure:

**Title:**
Assessment of the carrying value of retail design center long-lived assets

**Description:**
As discussed in Note 3 to the consolidated financial statements, the Company reviews long-lived assets for impairment whenever events or changes in circumstances indicate that the carrying value of these assets may not be recoverable. If the sum of the estimated undiscounted future cash flows over the remaining life of the primary asset is less than the carrying value, the Company recognizes a loss equal to the difference between the carrying value and the fair value. As of June 30, 2021, property, plant, and equipment, net, was $231.4 million, and the Company recognized impairment charges of $0.6 million for the fiscal year ended June 30, 2021. We identified the assessment of the carrying value of retail design center long-lived assets as a critical audit matter. Specifically, complex auditor judgment was required to assess the sales growth rates used to estimate the forecasted cash flows as they involved a high degree of subjectivity.

**Procedure:**
The following are the primary procedures we performed to address this critical audit matter. We evaluated the design and tested the operating effectiveness of certain internal controls over the Company's retail design center impairment assessment process, including controls related to the development of the sales growth rates. We evaluated the Company's sales growth rates by (1) comparing them to historical results, the Company's future operating plans, existing retail orders backlog, and industry reports, and (2) performing sensitivity analyses.

### Panel B: Structured format with component headers

Here is the breakdown of the critical audit matter regarding **Revenue Recognition:**

**Title:**
Revenue Recognition

**Description:**
The Company's revenue for the year ended January 31, 2022, totaled $281.4 million, primarily generated from cloud-hosted subscription fees, as described in Note 2 to the consolidated financial statements. To recognize revenue, the Company must evaluate whether promises made to customers represent distinct performance obligations, assess the appropriate measure of the transfer of control, and determine when the transfer of control has occurred. These assessments often require significant judgment, especially in contracts with non-standard terms. The audit was complex due to these non-standard contractual terms, which required significant auditor judgment to determine if distinct performance obligations were created. Proper identification of these performance obligations could significantly affect the timing of revenue recognition and related disclosures.

**Procedure:**
To address this matter in our audit, we obtained an understanding of the Company's process, evaluated the design, and tested the operating effectiveness of controls related to the identification and evaluation of performance obligations. This included the identification and consideration of non-standard contractual terms, the transaction price, and the measure of progress for the transfer of control. Our audit procedures involved reading a sample of contracts to evaluate whether management appropriately identified terms affecting revenue recognition and tested the Company's evaluation of standalone selling prices for its performance obligations. We also assessed the completeness and accuracy of the underlying data used in management's determination of standalone selling price, as well as the recorded deferred revenue and revenue amounts.

**Fig. A-3 Flowchart of the framework**

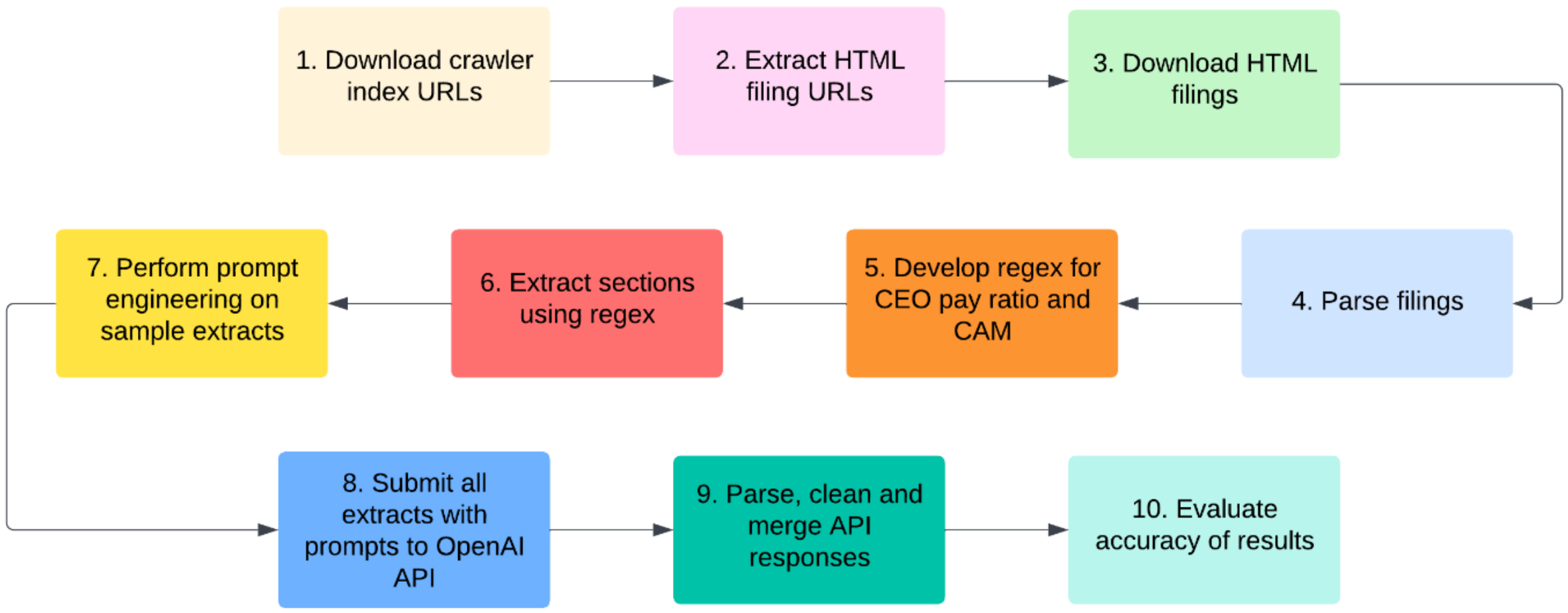